\documentclass[%
 reprint,
amsmath,amssymb,
aps,
]{revtex4-2}

\usepackage[utf8]{inputenc}
\usepackage[table]{xcolor}
\usepackage{graphicx}
\usepackage{dcolumn}
\usepackage{bm}
\usepackage{siunitx}
\usepackage{gensymb}
\usepackage{chemformula}
\usepackage[T1]{fontenc}
\usepackage{txfonts}
\usepackage{amsmath}
\usepackage{float}
\usepackage{hyperref}
\usepackage{changes}
\usepackage{braket}
\usepackage{physics}
\usepackage{multirow}

\begin{document}

\preprint{APS/123-QED}


\title{Spin qubit operations by conveyor-mode shuttling}
\author{M. De Smet$^{1}$}
\thanks{These authors contributed equally}
\author{Y. Matsumoto$^{1}$}
\thanks{These authors contributed equally}
\author{D. Fernández-Fernández$^{2, 4}$}
\author{L. Tryputen$^{3}$}
\author{S.L. de Snoo$^{1}$}
\author{D.J. Michalak$^{3}$}
\author{H.G.J. Eenink$^{3}$}
\author{G. Platero$^{2, 5}$}
\author{S. Bosco$^{1}$}
\author{G. Scappucci$^{1}$}
\author{L.M.K. Vandersypen$^{1}$}
\email{L.M.K.Vandersypen@tudelft.nl}

\affiliation{$^{1}$QuTech and Kavli Institute of Nanoscience, Delft University of Technology, Lorentzweg 1, 2628 CJ Delft, The Netherlands \\ $^{2}$Instituto de Ciencia de Materiales de Madrid ICMM, CSIC, Madrid 28049, Spain
\\ $^{3}$Netherlands Organization for Applied Scientific Research (TNO), Stieltjesweg 1, 2628 CK Delft, The Netherlands
\\ $^{4}$Institute of Physics, University of Augsburg, Augsburg, Germany
\\ $^{5}$Quantum Advanced Research Center (QuARC), CSIC, Madrid 28049, Spain
}

\date{\today}

\begin{abstract}

Dynamic qubit routing is emerging as a promising architectural path for semiconductor quantum processors. Charge carriers can be rapidly moved around on a chip using traveling-wave potentials known as conveyors, preserving the spin state with high fidelity. Originally developed for spin transport, conveyor-mode shuttling may also offer opportunities for performing qubit operations directly controlled by the motion itself.
Here, we demonstrate coherent single- and two-qubit control by conveyor-mode electron shuttling, using two conceptually different approaches. First, conveyor electric-dipole spin resonance (conveyor EDSR) achieves high-fidelity rotations by resonantly shuttling spins through transverse magnetic-field gradients at their mean Larmor frequency. Second, conveyor diabatic gates exploit quantization-axis tilts for tunable bang–bang control. Combining diabatic conveyor transport with exchange activation controlled by the motion directly yields a variety of effective two-qubit interactions selectable via the shuttling speed and distance. These experimental results motivate an architectural paradigm of reconfigurable and transport-driven spin qubits.

\end{abstract}

\maketitle

\section{\label{sec:Introduction}Introduction}

Semiconductor spin qubits have emerged as one of the leading platforms in the pursuit of scalable quantum computing, offering promising prospects for high qubit densities and integration with mature semiconductor manufacturing~\cite{vandersypen_quantum_2019, maurand_cmos_2016, zwerver_qubits_2022, steinacker_industry-compatible_2025, george_12-spin-qubit_2025, neyens_probing_2024}. Recent experimental advances include single-~\cite{yoneda_quantum-dot_2018, yang_silicon_2019, lawrie_simultaneous_2023, wu_simultaneous_2025, steinacker_industry-compatible_2025,team_digitally_2026,madzik_operating_2025} and two-qubit~\cite{xue_quantum_2022, noiri_fast_2022, mills_two-qubit_2022, tanttu_assessment_2024,team_digitally_2026} gate fidelities well above $99\%$. 
In these experiments, the spins remain localized in static quantum dots, and two-qubit interactions are constrained to nearest neighbors only. A common strategy for scaling from present-day experimental demonstrations to millions of qubits rests on tiling fixed-size qubit registers, connected together by on-chip quantum links~\cite{vandersypen_interfacing_2017}. Recent works have demonstrated the promise of realizing quantum links by physically displacing (shuttling) electrons across the chip while preserving their spin state. Conveyor-mode shuttling in particular, whereby phase-shifted sinusoidal voltages create a traveling-wave potential~\cite{taylor_fault-tolerant_2005, seidler_conveyor-mode_2022, xue_sisige_2024, struck_spin-epr-pair_2024, ademi_distributing_2025}, has allowed coherent spin transport over an effective distance of 10 micrometers with 99.5\% fidelity in under 200 ns~\cite{de_smet_high-fidelity_2025}.

Building on these advances, spin shuttling is now viewed not only as a means to connect separate qubit registers but also as a method for dynamically reconfiguring the qubit connectivity, as in a recent realization of four-way parity measurements~\cite{undseth_weight-four_2026}. Thinking one step further, one could imagine that static dots are abandoned altogether by performing quantum gates directly on the spins inside the conveyor potentials~\cite{cai_looped_2023,kunne_spinbus_2024}. As a case in point, a recent experiment demonstrated that a high-fidelity two-qubit gate could be realized directly on mobile spins, by moving two electrons in separate conveyor minima towards each other for a calibrated amount of time~\cite{matsumoto_two-qubit_2025}. The question arises whether also single-spin manipulation {\em controlled by the motion itself} might be feasible, exploiting spin-orbit coupling to generate spin rotations~\cite{fernandez-fernandez_flying_2024, bosco_high_2024}, analogously to early experiments in semiconductor spintronics~\cite{wolf_spintronics_2001}.

In this work, we introduce and demonstrate several concepts for the control of spin qubits by shuttling. First, we test conveyor electric-dipole spin resonance (EDSR)~\cite{kunne_spinbus_2024, pazhedath_large_2025}, where spin rotations are obtained by shuttling back-and-forth through a magnetic field gradient at a shuttling frequency resonant with the Larmor frequency.
This extends the principle of micromagnet EDSR by replacing the small electrically driven orbital motion of an electron in a static quantum dot~\cite{golovach_electric-dipole-induced_2006} with controlled shuttling over much larger distances, allowing the spin to sample substantially larger magnetic-field variations. Like flopping-mode EDSR~\cite{croot_flopping-mode_2020}, conveyor EDSR aims to increase the spin driving speed, but does so while avoiding charge delocalization between two static dots.

Second, we explore spin-diabatic conveyor shuttling, in which spin rotations emerge from diabatic motion through a spatially varying magnetic field. As the electron is transported along the channel, the direction of its spin quantization axis changes. For slow motion, the spin adiabatically follows this axis. In contrast, fast transport results in coherent precession about the new quantization axis. Such diabatic gates have previously been realized by tunneling between adjacent dots~\cite{wang_operating_2024, unseld_baseband_2025}. In conveyor-mode shuttling, natural and robust control of the rotation angle could potentially be achieved through either the shuttling speed~\cite{ademi_distributing_2025} or the shuttling distance, which provides a high degree of operational flexibility.

Third, we exploit quantization-axis tilts to directly implement different types of two-qubit gates controlled by the shuttling distance and speed, as was recently proposed~\cite{fernandez-fernandez_spin-orbit-enabled_2025}. Conventionally, when exchange is switched on rapidly and dominates over the difference in the qubit energy splittings, SWAP-like oscillations are produced. If the Zeeman energy dominates over the exchange coupling or if exchange is turned on slowly, CPhase oscillations emerge~\cite{meunier_efficient_2011,burkard_semiconductor_2023}. Conveyor-mode shuttling may additionally allow the direct implementation of controlled-unitary (CU-type) gates, where the rotation axis and angle of U are controlled by the micromagnet gradient and conveyor speed and distance.

\begin{figure}[t]
\includegraphics[width=0.97\columnwidth]{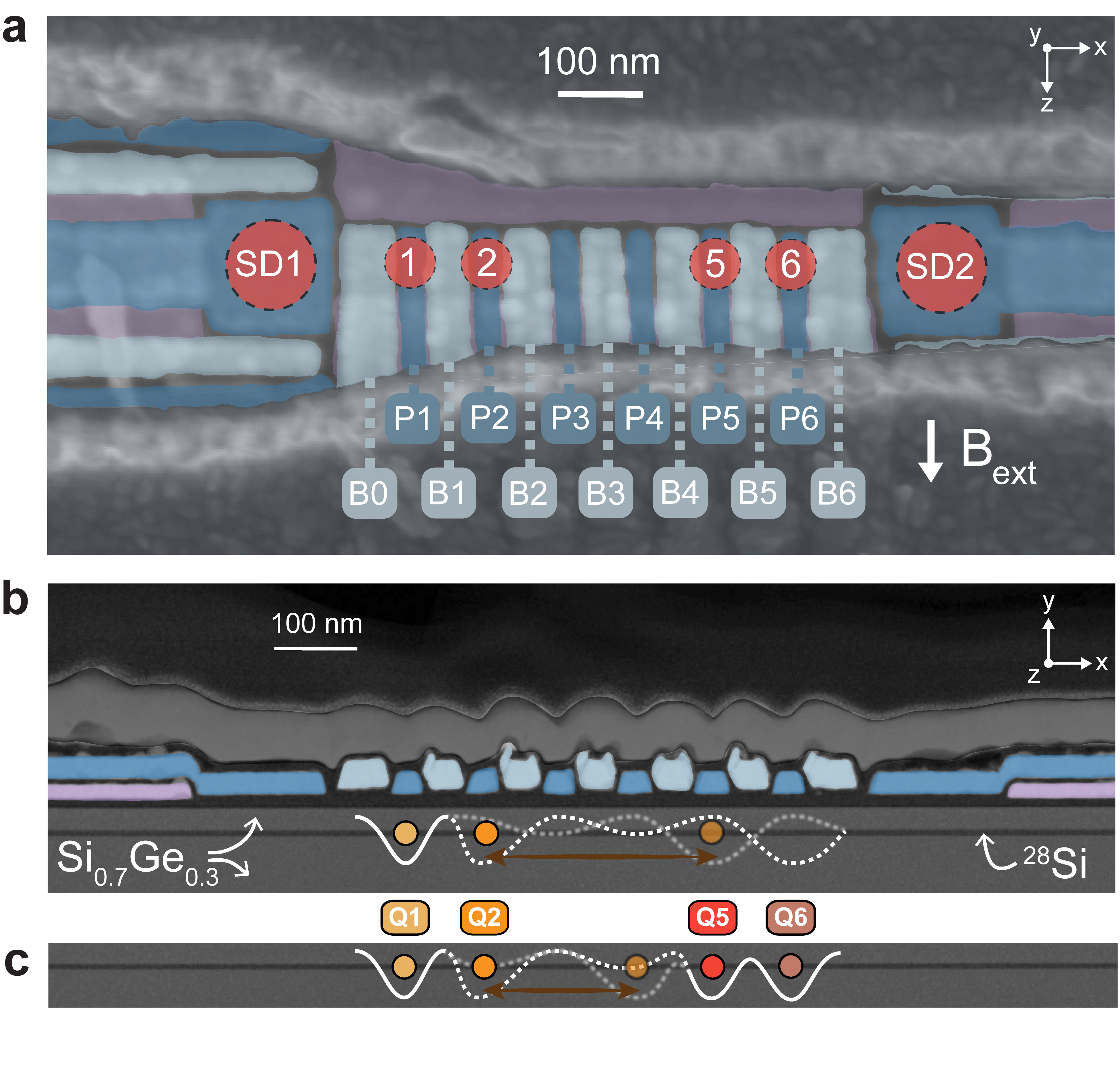}
\caption{\label{fig:fig1} \textbf{Spin qubit device.}
a) False-colored scanning electron microscope image and b) transmission electron microscope image of a nominally identical device to the one used in this work. The colors indicate different metallization layers. Plunger gates P1 through P6 (blue), barrier gates B0 to B6 (light blue) and two screening gates (purple) form a linear array of six quantum dots (where the static dots 1-2 and 5-6 are indicated by numbered circles) in a $\mathrm{^{28}Si/Si_{0.7}Ge_{0.3}}$ heterostructure. Two sensing dots, SD1 and SD2, are placed at the ends of the array. A cobalt micromagnet, seen in dark gray in a), is placed on top of the active area. The gray layer on top of the gate stack in b) is a Pt cap for improved imaging. Solid white lines at the Si quantum well represent the potential for static quantum dot 1, while the dotted lines represent the changing conveyor potential to transport qubit 2. In this condition, dots 5 and 6 are emptied for conveyor EDSR experiments. c) The second charge filling condition used in this work. Dots 5 and 6 are filled to the (3,3) occupation for diabatic shuttling of Q2 and the interaction with Q5. Qubits Q1 and Q6 are used as ancillas for readout.}
\end{figure}

\section{\label{sec:Results}Results}

\begin{figure*}[t]
\includegraphics[width=\textwidth]{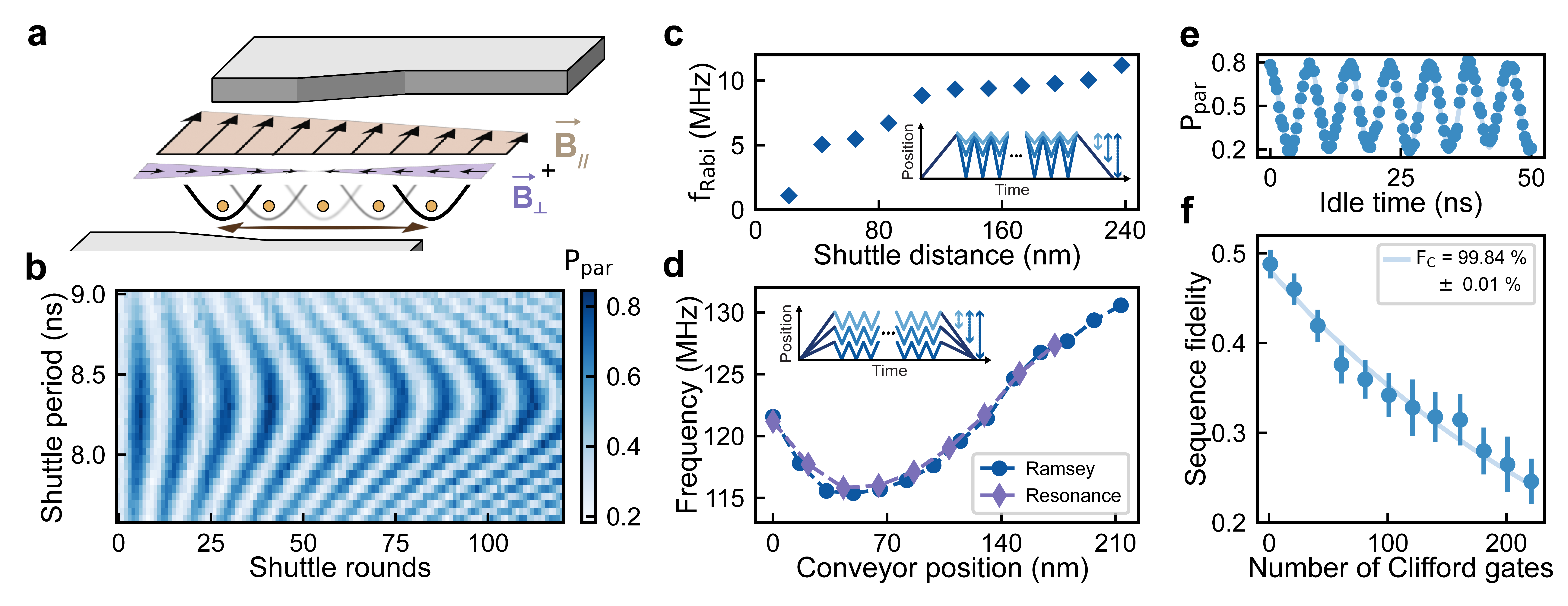}
\caption{\label{fig:fig2} \textbf{Conveyor EDSR}.
a) Schematic depiction of back-and-forth conveyor shuttling of a spin through magnetic field gradients. Here $\overrightarrow{B}_{\parallel}$ ($\overrightarrow{B}_{\perp}$) is the local field component parallel (perpendicular) to the average magnetic field direction along the shuttled path. b) Conveyor EDSR chevron pattern obtained by shuttling a single spin back and forth over a fixed distance of 219 nm for a variable number of round trips and time per trip (shuttle period). The spin qubit (Q2) is first shuttled from quantum dot 2 to a position around gate P5 in 8 ns, after which barrier B1 is closed more to avoid possible interaction with Q1 during back-and-forth shuttling. c) Rabi driving frequency dependence of conveyor EDSR when increasing the back-and-forth shuttling distance. The electron starts under gate P5 and is shuttled towards gate P2. d) Conveyor EDSR for a back-and-forth shuttling distance of approximately 43 nm performed at different positions along the conveyor channel. The conveyor EDSR resonance frequency matches the mean local Larmor frequency probed with Ramsey measurements very well. Here, the zero position is defined as the start of the conveyor trajectory, 54 nm to the right of the center of gate P2. e) Ramsey oscillations of Q2 when idling at a position of 270 nm, corresponding to the middle of gate P5. Here, wait times are used to implement Z-control of the qubit.  f) Randomized benchmarking (RB) of conveyor EDSR (see Methods for the gateset), yielding a Clifford gate fidelity of $99.84 \pm 0.01\%$. The sequence fidelity represents the difference between RB measurements with input states $\ket{\uparrow}$ and $\ket{\downarrow}$, each with 50 randomizations. The error bars correspond to $95\%$ confidence intervals, and the uncertainty in the fidelity is one standard deviation extracted from the exponential fit.}
\end{figure*}

\subsection{\label{subsec:Device}Device design and operation regime}

The spin qubit device is able to host a linear array of six quantum dots (Fig.~\ref{fig:fig1}a), fabricated on a $^{28}$Si/SiGe heterostructure (Fig.~\ref{fig:fig1}b). A cobalt micromagnet on top of the gate electrodes enables EDSR for single-qubit rotations~\cite{obata_coherent_2010} and creates position-dependent spin resonance frequencies for the electrons. The device is identical to that used in~\cite{de_smet_high-fidelity_2025, matsumoto_two-qubit_2025}. We magnetize the micromagnet with an external magnetic field $B_{ext}$ of either 1.5 T or 2 T, after which we lower $B_{ext}$ to only a few mT. In the course of the experiments, we additionally sweep the magnetic field up and down over small field ranges (up to about 60 mT) in order to produce the desired total magnetic field strength and gradients for any given experiment (see Supplementary Information~\ref{supp:subsec:demagnetization}).

We form quantum dots (QDs) 1 and 2 and operate in the (3,1) electron occupation. Here we initialize in the $\ket{\downarrow \uparrow}$ state~\cite{petta_coherent_2005} by ramping the detuning from the (4,0) occupation and read out using parity Pauli Spin Blockade (PSB)~\cite{ono_current_2002, johnson_singlet-triplet_2005}. The initialization sequence includes an initial readout following the detuning ramp, enabling post-selection of the anti-parallel spin states to enhance the initialization fidelity. In initial experiments, we empty all other QDs, including 5 and 6. The conveyor is initialized by pulsing the gate voltage configuration such that the electron in dot 2 is transferred to the first conveyor minimum, located 54 nm to the right of the center of gate P2. We then shuttle that electron spin up to an additional 216 nm, for a total distance from P2 of 270 nm, reaching the position of P5. When studying two-qubit gates, we similarly load a spin from dot 2 into the conveyor and operate QD pair 5-6 in the (3,3) charge state, ramping adiabatically through the ST-anticrossing to initialize in the $\ket{\downarrow \downarrow}$ state.

\subsection{\label{subsec:Conveyor EDSR}Conveyor EDSR}

For conveyor EDSR, we shuttle an electron spin back and forth, such that it experiences a periodically varying transverse magnetic field gradient from the micromagnet stray field (schematically illustrated in Fig.~\ref{fig:fig2}a). When this back-and-forth shuttling frequency matches the mean Larmor frequency along the shuttled path, the resonance condition is met, and coherent spin rotations are induced. The mean Larmor frequency, therefore, needs to fall in the range of accessible back-and-forth shuttling periods, given the maximum available conveyor speed of 64 m/s~\cite{de_smet_high-fidelity_2025}. At an external magnetic field of -8 mT (see Supplementary Information~\ref{supp:subsec:demagnetization}), we obtain Larmor frequencies between 115 and 130 MHz. Under this condition, we coherently drive the spin via conveyor EDSR, revealing a well-defined chevron pattern (see Fig.~\ref{fig:fig2}b). However, different from a typical chevron with a continuous driving time axis, here the driving is expressed in terms of integer numbers of shuttle rounds. The vertical axis shows the shuttle period instead of the usual drive frequency, resulting in an asymmetric chevron pattern. Keeping the shuttle distance fixed at 219 nm, we find that resonance occurs at a shuttle period of 8.32 nm, corresponding to a mean Larmor frequency of 120.2 MHz. Although the measurements in the main text employ a triangular shuttling trajectory, conveyor EDSR can also be realized using alternative motion profiles, including sinusoidal and block-like trajectories (Supplementary Information~\ref{supp:subsec:mod_CV}). Resonance also occurs at subharmonics of the mean Larmor frequency (Supplementary Information~\ref{supp:subsec:subharmonics}).

Figure \ref{fig:fig2}c shows that the Rabi frequency $f_{Rabi}$ increases with the range of the back-and-forth motion, reflecting the larger transverse magnetic-field difference the spin experiences. We speculate that a spatially varying gradient is responsible for the non-linear trend in $f_{Rabi}$. Notably, the Rabi decay time \(T_2^{Rabi}\) (Fig.~\ref{supp:fig:Rabi_decay}) improves with increased shuttling distances (see Supplementary Information~\ref{supp:subsec:Rabi_decay}). This is consistent with the Rabi decay being dominated by qubit frequency fluctuations, with motional narrowing~\cite{mortemousque_enhanced_2021, langrock_blueprint_2023, struck_spin-epr-pair_2024, krzywda_coherence_2026}, and with the dependence of spin coherence on position (see Supplementary Information~\ref{supp:subsec:T2*}). As both the Rabi frequency and Rabi decay time increase with shuttle distance, the quality factor varies over the explored distance range by up to a factor of 58.

To further validate the resonant driving mechanism underlying conveyor EDSR, we perform additional measurements using smaller shuttling segments of 43 nm at various positions. The resonance frequencies extracted from these local scans (Fig.~\ref{fig:fig2}d) closely match the local Larmor frequencies determined from static-conveyor Ramsey experiments, confirming that the drive mechanism operates as expected.

Finally, we benchmark the conveyor EDSR performance, where conveyor EDSR rotations are defined to be around the $x$-axis, whereas $z$-axis control is achieved through Larmor precession with the spin confined in a static conveyor potential near gate P5 (Fig.~\ref{fig:fig2}e) \cite{wang_operating_2024, unseld_baseband_2025}. Setting the conveyor distance to 221 nm yields a Rabi frequency corresponding to an X$_{90}$ gate executed in exactly three shuttle rounds. We construct a gateset based on the native positive X\(_{90}\) and Z-rotation gates (see Methods) to perform randomized benchmarking. The resulting Clifford gate fidelity reaches 99.84\%, with most of the error originating from the X\(_{90}\) gates (see interleaved randomized benchmarking results in Supplementary Information~\ref{supp:subsec:IRB} and Supplementary Information~\ref{supp:subsec:read_wait}). Given that there are two X\(_{90}\) operations per Clifford gate, the extracted X\(_{90}\) gate fidelity \(F_{X_{90}}\) exceeds 99.9\%. We speculate that the measured X\(_{90}\) gate infidelity is in part due to a small unintentional drive of the Q1 ancilla spin, with which we compare the state of the shuttled spin, so the actual X\(_{90}\) fidelity of the target spin may be higher. 

\subsection{\label{subsec:diab}Diabatic conveyor gate}

\begin{figure*}[t]
\includegraphics[width=\textwidth]{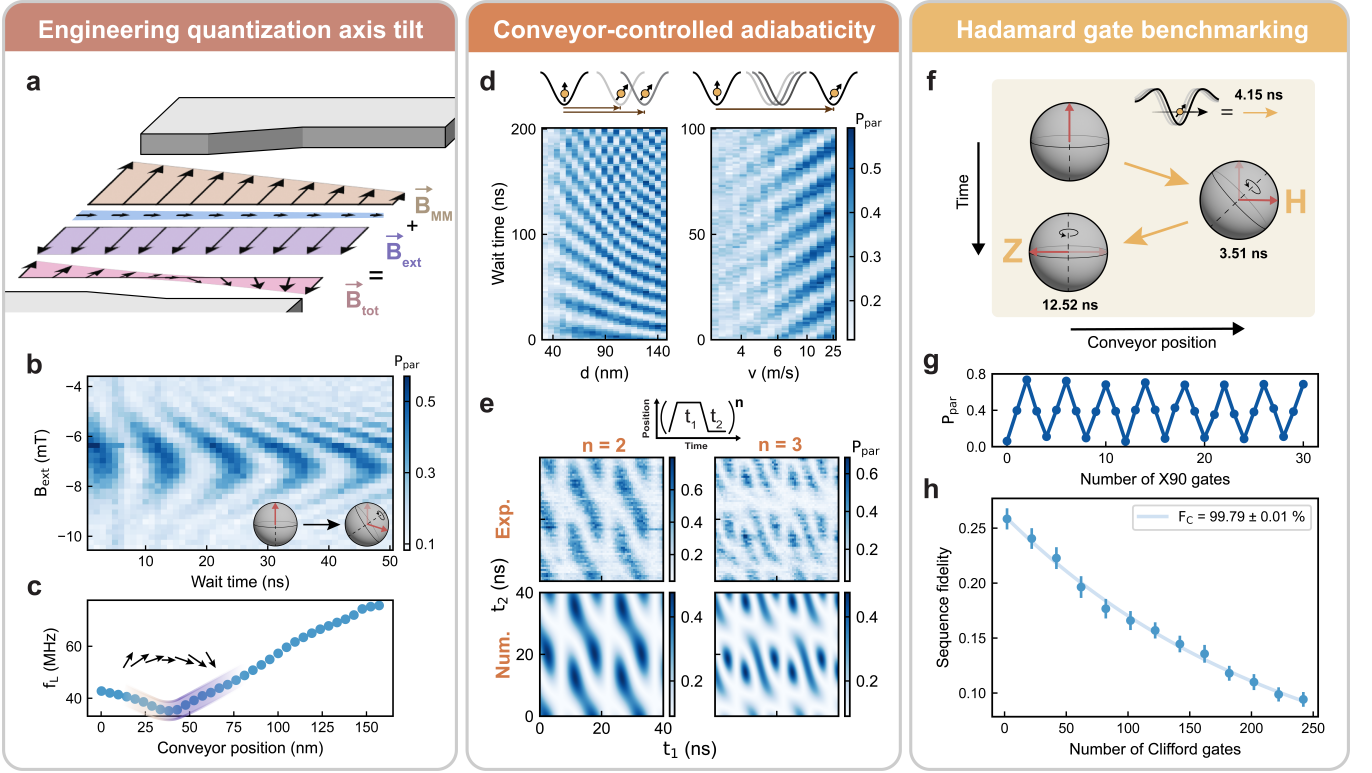}
\caption{\label{fig:fig3} \textbf{Diabatic conveyor shuttling}.
a) Schematic of conveyor shuttling through a region with a strong tilt in the total magnetic field $\overrightarrow{B}_{tot}$ direction, engineered by an external magnetic field $\overrightarrow{B}_{ext}$ opposing the micromagnet (gray) stray field $\overrightarrow{B}_{MM}$. A small transverse field component (blue) is introduced by the micromagnet stray field and external field misalignment (see Supplementary Information~\ref{supp:subsec:demagnetization}). b) Rotations of Q2 around its new quantization axis as a function of time after shuttling over 139.6 nm at a speed of 53.7 m/s. $B_{ext}$ controls the Larmor frequency and tilt angle along the shuttled path, which in turn governs the degree of diabaticity of spin shuttling. The micromagnet magnetization was changed in subsequent experiments. c) At $B_{ext}$ = -3.5 mT, the Larmor frequency shows a minimum around 40 nm. Here, the strongest tilt in the quantization axis can be found. d) The spin-shuttling degree of diabaticity depends on both the shuttled distance and speed. The left plot indicates that diabatic rotations only occur after traversing the strong tilt location at 40 nm. The right plot shows that for a fixed distance of 129.6 nm, the shuttle speed effectively controls the degree of diabaticity. e) When shuttling for $n$ round trips, with variable idle times $t_1$ and $t_2$ at the final and initial positions, 129.6 nm and 0 nm respectively, interference patterns in the spin polarization appear. The top plots show experimental data for $n = 2$ and $n=3$, while the bottom plots show the numerically simulated patterns. Here, the final position of Q2 is 129.6 nm, and the conveyor speed is 16.2 m/s, such that the one-way shuttling time is $t_s$ = 8 ns. f) The conveyor speed and distance can be tuned to the semi-diabatic regime such that the resulting effective quantization-axis tilt is exactly 45 degrees. By adjusting the idling time at the final position, a Hadamard (H) gate can be executed. Arbitrary single-qubit gates can be implemented by combining Z gates in the initial position with H gates in the final position. g) Repeated X\(_{90}\) gates showing a discretized version of Rabi oscillations, where we implemented X\(_{90}\) as H followed by Z$_{180}$. h) Randomized benchmarking of the spin-diabatic conveyor gate, yielding a Clifford fidelity of $99.79 \pm 0.01\%$. The sequence fidelity represents the difference between RB measurements with input states $\ket{\uparrow}$ and $\ket{\downarrow}$. The data point error bars correspond to $95\%$ confidence intervals, the uncertainty in the fidelity is one standard deviation extracted from the exponential fit.}
\end{figure*}

While conveyor EDSR achieves high-fidelity operation, conveyor-mode shuttling can also be used to explore diabatic spin control, suddenly displacing the spin to a position where the quantization axis is tilted relative to the starting position. This approach does not rely on resonant control and can possibly enable even faster gate operations.

\begin{figure*}[!t]
\includegraphics[width=\textwidth]{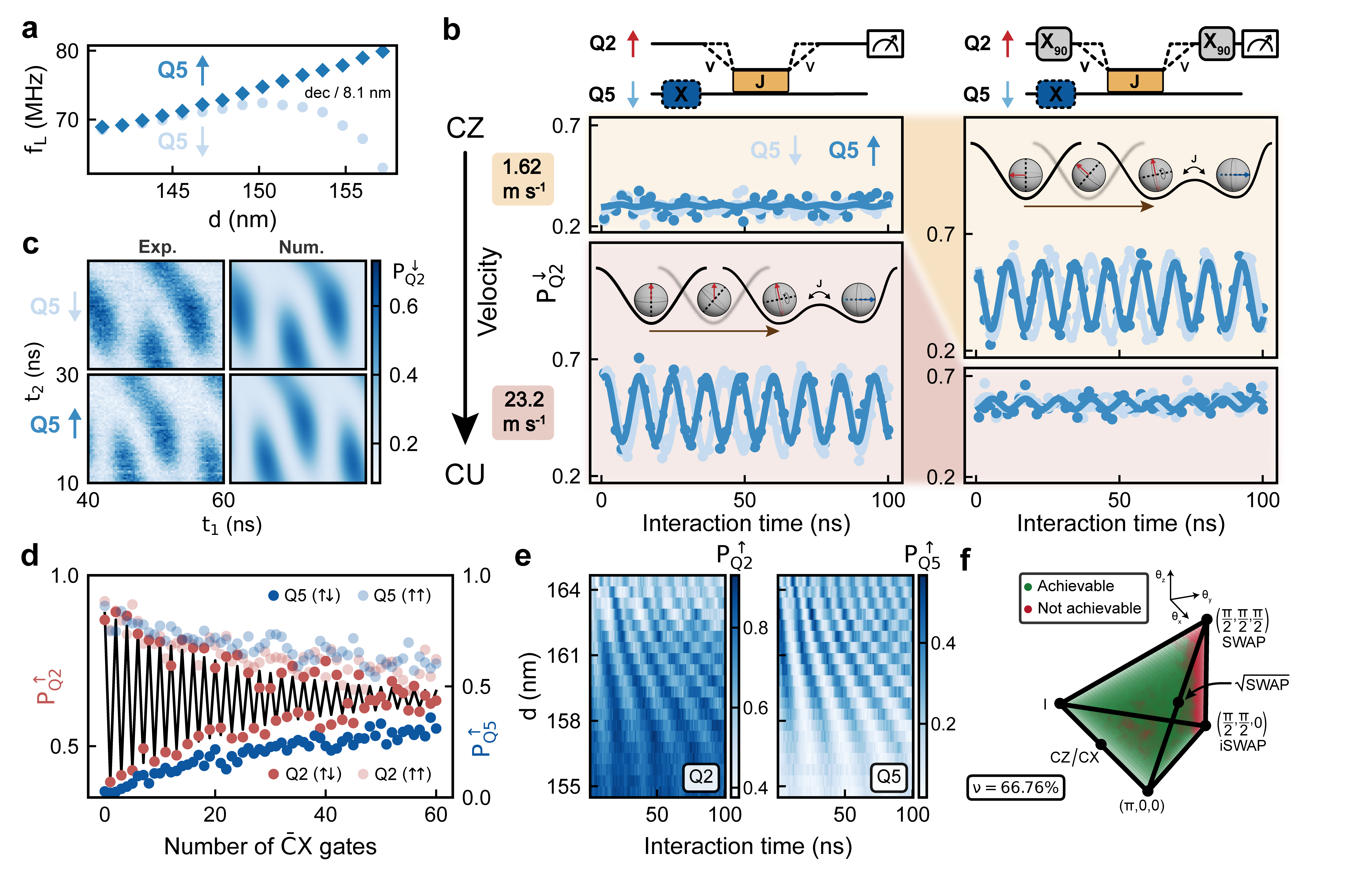}
\centering
\caption{\label{fig:fig4} \textbf{Two-qubit gates with diabatic spin shuttling}. a) Larmor frequency of Q2 as a function of the shuttled distance $d$, extracted from diabatic shuttling oscillations similar to Fig.~\ref{fig:fig3}d. When Q2 approaches Q5, the exchange coupling is activated with one decade per 8.1 nm. The Larmor frequency of Q2 now depends on the spin state of Q5. b) Spin-down probability after single-round shuttling of Q2, which is initialized either in $\ket{\uparrow}$ or a superposition state (columns), over a distance of 155 nm for two values of the shuttle velocity $v$ (rows). At low velocity, Q2 is shuttled adiabatically with respect to the quantization-axis tilt and performs a CPhase gate with Q5 after 40.1 ns. This is evident only when initializing and measuring Q2 in the X basis (top right panel). At high velocity, Q2 is shuttled diabatically with respect to the quantization-axis tilt. For a 90 degree tilt, this would effectively interchange the X and Z bases, and a CZ gate would transform into a CX gate. For general tilt angles, it becomes a controlled-unitary rotation CU. c) The interference patterns for two-round shuttling with variable idle times, see the left plots of Fig.~\ref{fig:fig3}e, now become conditioned on the Q5 spin state. The conveyor speed is 16.2 m/s, and waiting time $t_1$ ($t_2$) occurs at 153.9 nm (0 nm). The numerical fits on the right side allow us to extract the Zeeman splittings, tilt angle, and exchange strength (Supplementary Information~\ref{supp:subsec:diabatic_analysis}). d) Repeated $\mathrm{\bar{C}X}$ gates with Q2 as the target and Q5 as the control qubit, showing the spin-up probabilities $P^{\uparrow}_{Q2}$ and $P^{\uparrow}_{Q5}$. The legend labels the datapoints for Q2-Q5 initialized in either the $\ket{\uparrow\downarrow}$ or the $\ket{\uparrow\uparrow}$ state. The micromagnet condition did not produce a tilt angle of 90 degrees or above for Q2. Therefore, to achieve a $\mathrm{\bar{C}X}$ (or CX) gate, we shuttle over two rounds with identical shuttle and interaction conditions and add an intermediate Z rotation of Q2 in a static dot below gate P2. The total time per $\mathrm{\bar{C}X}$ gate is 157.1 ns, with 50.5 ns between repeated gates. e) SWAP-type oscillations between Q2 and Q5, showing the respective spin-up probabilities. The spins are initialized in the $\ket{\uparrow \downarrow}$ state, Q2 is shuttled adiabatically with respect to the quantization axis change but diabatically with respect to the exchange rate of change. Further analysis can be found in Supplementary Information~\ref{supp:subsec:ST_osc}, where we identify a shuttling-induced $ST_{-}$ anti-crossing of 34.6 MHz. f) A theoretical estimate of the 2Q gates achievable with unitary error below 0.1\%, visualized in the Weyl chamber. The estimate covers 66.76\% of the chamber volume and assumes a minimum 12.8 degree quantization-axis tilt between the interacting two spins (see Supplementary Information~\ref{supp:subsec:diabatic_analysis}).
}
\end{figure*}

A small external magnetic field is applied opposite in direction but comparable in magnitude to the micromagnet stray field in the shuttling channel, engineering a strong spatial variation in the total magnetic field direction, as schematically illustrated in Fig.~\ref{fig:fig3}a. The critical role of the external magnetic field compensating the stray field is clearly seen in Fig.~\ref{fig:fig3}b, where diabatic single-spin rotations are observed in a narrow magnetic field range only. In addition to a sufficient tilt, we want the Larmor frequencies to be low enough to achieve good timing control of the rotations and a large degree of diabaticity when shuttling. We therefore demagnetize the micromagnet further compared to the conveyor EDSR regime. Ramsey measurements along different locations of the conveyor channel map the spatial variation in Larmor frequency. The observed kink in Fig.~\ref{fig:fig3}c, around 40 nm from the start of the conveyor, indicates the location of the strong quantization-axis tilt (see Supplementary Section~\ref{supp:subsec:demagnetization}).

Fig.~\ref{fig:fig3}d clearly shows that shuttling-induced Larmor precession around a new quantization axis appears only when the electron is displaced beyond 40 nm, as expected. Furthermore, we see that the contrast of the oscillations vanishes if the conveyor velocity is reduced. The qubit dynamics thus depends not only on the shuttling distance but also on the shuttle velocity, which in the present experiment can be tuned to access regimes ranging from fully adiabatic to semi-diabatic transport. The shuttle distance (strictly speaking the start and end points) and speed thus enable control over the effective rotation axis. To probe the tilt of the quantization axis for a specific shuttling distance and velocity, we shuttle the electron back and forth multiple times between two positions, allowing the spin to precess around the local quantization axes. By varying the wait times at these positions, we observe a characteristic interference pattern that allows numerical reconstruction of the unitary evolution of the spin, following the approach of \cite{wang_operating_2024, unseld_baseband_2025}. Fig.~\ref{fig:fig3}e exemplifies how we extract a tilt angle of 68.4 degrees for a shuttled distance of 129.6 nm at a speed of 16.2 m/s.

For a shuttled distance of 70.2 nm, tuning the conveyor speed to 16.9 m/s places the system in the semi-diabatic regime where the effective quantization-axis tilt reaches exactly 45 degrees. When the elapsed time at that position corresponds to one half of a Larmor precession, shuttling back and forth just once then realizes a Hadamard gate H. An additional idle time back at the initial position implements a Z$_{180}$ rotation, which, when combined with the H, yields an effective X$_{90}$ gate in just one shuttle round (Fig.~\ref{fig:fig3}f). The $X_{90}$ gate is calibrated through AllXY measurements and by repeated gate application, as shown in Fig.~\ref{fig:fig3}g. Figure~\ref{fig:fig3}h shows randomized benchmarking of this (semi-)diabatic conveyor gate, using Clifford gates constructed from \{X$_{90}$,\, Z$_{180}$,\, Z$_{270}$,\, Z$_{450}$\}, such that each Clifford gate in this set contains exactly two X$_{90}$ operations (see Methods). This yields a single-qubit Clifford gate fidelity of 99.79\%.

\subsection{\label{subsec:2Q}Two-qubit interaction types}

We next load spin qubits into quantum dots 5 and 6, and combine the shuttling of Q2 through a region of strong quantization-axis tilt with distance-tunable exchange coupling to Q5 \cite{matsumoto_two-qubit_2025}. When Q2 is shuttled diabatically towards stationary Q5, the exchange interaction is activated, and the precession frequency of Q2 around its new quantization axis becomes conditional on the spin state of Q5 (Fig.~\ref{fig:fig4}a) along its own quantization axis (see Supplementary Information~\ref{supp:subsec:diabatic_analysis}).

As shown in Fig.~\ref{fig:fig4}b, when the exchange interaction remains below the Zeeman splitting difference between the qubits, spin-adiabatic shuttling of Q2 leads to a controlled-Z (CZ) interaction up to single-qubit phase corrections. Increasing the shuttling speed effectively updates the frame of Q2, transforming the conditional phase rotations into controlled-unitary operations. Here, CU denotes rotations of Q2 around a tilted quantization axis, where Q5 acts as the control qubit. When the quantization-axis tilt between the left and right shuttle positions reaches or exceeds 90 degrees, the CU operation can be tuned to a controlled-X (CX) gate.

Figure \ref{fig:fig4}c shows that the interference pattern of Q2 from Fig.~\ref{fig:fig3}e now depends on the state of Q5. These interference patterns provide a means of calibration for a CX (or the zero-controlled $\mathrm{\bar{C}X}$) operation in two shuttle rounds with an intermediate Z rotation. Moreover, fitting the experimental data with numerical simulations yields a quantization axis for Q5 that is tilted by 83.6 degrees compared to the zero position of Q2 (see Supplementary Information~\ref{supp:subsec:diabatic_analysis}).

Repeated applications of $\mathrm{\bar{C}X}$ gates with Q5 prepared in the spin-down or spin-up (flipped using conveyor EDSR, see Supplementary Information~\ref{supp:subsec:EDSR_Q5}) state are shown in Fig.~\ref{fig:fig4}d. Q2 is flipped only when Q5 is initialized to the spin-down state. The characteristic decay of the $\mathrm{\bar{C}X}$ operation is overshadowed by gradual randomization of Q5, also visible in the figure. This effect may originate from off-resonant driving of Q5 or Q6, or from readout quality degradation with increasing gate applications, which impacts the reliability of post-selection used for Q5 initialization. As a result, imperfect state preparation of Q5 increasingly mixes the control states, accelerating the decay of the Q2 oscillations, hence obscuring the intrinsic $\mathrm{\bar{C}X}$ fidelity.

In Fig.~\ref{fig:fig4}d, the shuttling distance is chosen such that the exchange coupling remains relatively weak ($J$ = 4 MHz), well below the Zeeman energy difference (see Supplementary Information~\ref{supp:subsec:diabatic_analysis}), and the shuttle speed is 16.9 m/s. Extending the shuttling distance to enhance $J$ and shuttling at 1.93 m/s clearly reveals exchange-driven oscillations (Fig.~\ref{fig:fig4}e), with a maximum extracted frequency of 93.8 MHz, approaching the SWAP regime. This shuttle speed corresponds to adiabatic motion with respect to the changing magnetic-field vector but diabatic dynamics with respect to the change in $J$ given $\Delta E_Z$.

The Weyl chamber~\cite{zhang_geometric_2003,tucci_introduction_2005,calderon-vargas_directly_2015} in Fig.~\ref{fig:fig4}f indicates the range of two-qubit gate types achievable in a single step in our system. Here, we define a two-qubit gate as achievable when the theoretical unitary error is below 0.1\%, and we assume that the quantization axis angle between Q5 and the zero position of Q2 is 83.6 degrees for the current two-qubit interaction zone. We find a Weyl chamber filling factor of 66.76\%. The filling factor increases with the misalignment of the qubit quantization axes. Therefore, more two-qubit gate types could be achieved in the current device by positioning the interaction zone exactly at the location of the strongest axis tilt, centered around 40 nm.

\subsection{\label{subsec:discussion}Discussion}

Conveyor EDSR and diabatic conveyor control achieve comparable gate fidelities, yet rely on fundamentally different control mechanisms. Conveyor EDSR employs frequency-selective resonant motion, while diabatic gates arise from non-adiabatic spin evolution during time-controlled motion between positions with different quantization axes. Because the shuttling time is finite, the control is not strictly bang–bang, and additional phase evolution occurs. However, this semi-diabatic evolution can be accurately calibrated and exploited. In turn, an $X_{90}$ using conveyor EDSR in just three shuttling rounds implies a Rabi frequency just twelve times lower than the Larmor frequency, such that the rotating-wave approximation is only partly justified. The resulting corrections are likewise absorbed in the calibrated gate operation. We also note that conveyor motion does not inherently induce qubit rotations. Coherent control arises only when the shuttling parameters are intentionally chosen to satisfy resonance conditions or to induce diabatic evolution. Otherwise, the spin can be transported while accumulating only a deterministic phase.

Neither shuttling control scheme provides native Y gates and thus requires physical Z rotations, though we note that negative X rotations can, in principle, be implemented in the conveyor EDSR scheme by reversing the initial shuttling direction. Z control is typically realized through idling-induced Larmor precession, but each idling qubit will evolve at its own Larmor frequency. Individual Z rotations could alternatively be implemented by shuttling spins through longitudinal magnetic-field gradients.

In terms of the gate duration, the diabatic X\(_{90}\) gate requires only a single shuttle round, whereas the EDSR implementation requires three. However, this advantage is largely offset by the slower Z rotations in the diabatic scheme, due to the lower Larmor frequencies used, leading to similar overall X$_{90}$ durations. Given the currently used shuttling speed and frequency, constrained by the arbitrary waveform generator bandwidth~\cite{de_smet_high-fidelity_2025}, it is natural to operate at low magnetic field, where Larmor frequencies are low. 

Finally, both conveyor EDSR and diabatic conveyor gates have their counterparts in bucket-brigade style shuttling, namely flopping-mode EDSR~\cite{benito_electric-field_2019, croot_flopping-mode_2020} and hopping gates~\cite{wang_operating_2024, unseld_baseband_2025} respectively. Conveyor EDSR improves on the flopping-mode principle from oscillation between two dots to continuous transport over larger distances. This approach samples larger magnetic-field gradients while avoiding the charge delocalization effects inherent to flopping-mode operation. 
For diabatic operation, the rotation is set by the tilt angles between the beginning and end points of the trajectory. Here, the larger distances traversed when using conveyors may simplify engineering deterministic tilt angles with micro- or nanomagnets, including tilt angles beyond 90 degrees. The hopping implementation is more constrained. Furthermore, a single conveyor could host multiple spins and enable a shared-control architecture, reducing the wire count and simplifying the fan-out~\cite{matsumoto_two-qubit_2025}. Selective control of individual spins can still be achieved sequentially through their shuttling trajectories as they pass local regions with micromagnet-engineered resonance frequencies for conveyor EDSR or strong quantization-axis tilts for diabatic conveyor gates.

\subsection{\label{subsec:conclusion}Conclusion}

This work demonstrates coherent single-spin manipulation using conveyor-mode shuttling, establishing a new paradigm for dynamic spin control in conveyor-based architectures. By synchronizing electron motion with the spin precession frequency, conveyor EDSR enables fast and high-fidelity single-qubit rotations without requiring strong charge delocalization or additional AC electric or magnetic fields at the qubit frequency. Extending beyond resonance-based control, spin-diabatic conveyor gates allow for bang-bang control when abruptly moving between two positions with distinct local quantization axes. Additionally, by reducing the conveyor velocity, lower effective tilts in the rotation axes can be obtained.

Combining the diabatic shuttling with conveyor-controlled exchange coupling to a stationary qubit, we have observed two-qubit time evolution under various effective coupling Hamiltonians selectable by the conveyor speed and distance traveled. Theoretically, more than half~\cite{calderon-vargas_directly_2015} of the two-qubit operations in the Weyl chamber can be realized in a single step, ranging from controlled-phase to controlled-X and SWAP-style oscillations. Our results establish conveyor-mode operation of single spins as a broad framework linking spin transport, local control, and entangling interactions within a single reconfigurable platform.

\section*{\label{sec:ackn}Acknowledgments}
We wish to thank Maximilian Rimbach Russ, Brennan Undseth, Lars Schreiber and the members of the Delft spin qubit groups for fruitful discussions. We are grateful to S.G.J. Philips for writing control libraries and designing the PCB, R. Schouten, R. Vermeulen, O. Benningshof and T. Orton for support with the measurement setup and dilution refrigerator. G.P. and D.F.F. are supported by the Spanish Ministry of Science through the grant: PID2023-149072NB-I00 and by the CSIC Research Platform PTI-001. D.F.F. acknowledges funding from the Emmy Noether Programme of the German Research Foundation (DFG) under grant no. BE 7683/1-1. We acknowledge financial support from the Army Research Office (ARO) under grant number W911NF-23-1-0110. The views and conclusions contained in this document are those of the authors and should not be interpreted as representing the official policies, either expressed or implied, of the ARO or the US Government. The US Government is authorized to reproduce and distribute reprints for government purposes notwithstanding any copyright notation herein. 

\section*{\label{sec:contrib}Author contributions}
M.D.S. and Y.M. performed the experiments and data analysis. D.F.-F. carried out simulation modeling of the diabatic shuttling gates, with supervision of G.P. and S.B. S.L.S. wrote software libraries for experimental control. M.D.S., Y.M. and L.M.K.V. contributed to data interpretation. L.T. fabricated the device. D.J.M. and H.G.J.E. managed the TEM study. G.S. provided the $^{28}$Si/SiGe heterostructure. M.D.S., Y.M. and L.M.K.V. wrote the manuscript with comments by all authors. M.D.S. and Y.M. conceived the experiments, and L.M.K.V. supervised the project.

\section*{\label{sec:interests}Competing interests}
M.D.S. and Y.M. are co-inventors of a patent application (NL2041283) concerning diabaticity control by conveyor shuttling. The other authors declare no competing interests.

\section*{\label{sec:repo}Data availability}
The raw measurement data and the analysis supporting the findings of this work are available in a Zenodo repository (https://doi.org/10.5281/zenodo.20413160).

\bibliographystyle{naturemag}

\bibliography{manualbib}

\section*{\label{sec:methods}Methods}

\subsubsection{Device fabrication}

The device is fabricated on an isotopically purified $^{28}$Si/SiGe heterostructure. A \SI{1.5}{\micro\meter} linearly graded \ch{Si_{1-x}Ge_x} buffer layer is grown on a silicon wafer, followed by a relaxed \SI{300}{\nano\meter} \ch{Si_{0.7}Ge_{0.3}} spacer. A tensile-strained \SI{7}{\nano\meter} \ch{^{28}Si} quantum well with a residual \ch{^{29}Si} concentration of 800 ppm is then grown~\cite{degli_esposti_low_2024}, separated from the gate stack by a \SI{30}{\nano\meter} \ch{Si_{0.7}Ge_{0.3}} spacer passivated with dichlorosilane at \SI{500}{\celsius}~\cite{degli_esposti_wafer-scale_2022}. Ohmic contacts to the two-dimensional electron gas are formed via phosphorus-ion implantation and electron beam evaporation of Ti:Pt with a thickness of 5:55 \SI{}{\nano\meter}.

The gate stack begins with a \SI{10}{\nano\meter} \ch{Al2O3} dielectric layer, above which three Ti:Pd gate layers of thickness 3:17, 3:27, and 3:37~\si{\nano\meter} are deposited by electron beam evaporation, each separated by \SI{5}{\nano\meter} \ch{Al2O3} grown by atomic layer deposition. These Ti:Pd layers define the screening, plunger (P), and barrier (B) gates that control the confinement potentials of the quantum dots. Sensing dots are placed at both ends of the linear array to enable charge sensing and to serve as electron reservoirs. A final \SI{5}{\nano\meter} \ch{Al2O3} capping layer is deposited over the gate stack, followed by electron beam evaporation of a 5:200~\si{\nano\meter} Ti:Co micromagnet used for qubit addressing and driving.

\subsubsection{Two-tone conveyor shuttling}

In the main text of this work, we employ a two-tone conveyor~\cite{de_smet_high-fidelity_2025} to shuttle qubit 2 or qubit 5. Each gate in the conveyor, therefore, receives two sinusoidal pulses, given by $V_n(t) = V^{DC}_n + \frac{A_n}{2}\big[\sin(2\pi ft - \phi_n) + \sin(\pi ft - \theta_n)\big]$. Here, the DC voltage $V^{DC}_n$ is tuned individually for each gate, while the amplitudes $A_n$ are shared across all gates within the same gate layer. The frequency $f$ and phases $\phi_n$ and $\theta_n$ are kept fixed and are not individually tuned. For conveyor EDSR, plunger and barrier gates received amplitudes of 130 mV and 160 mV, respectively. For diabatic conveyor shuttling, plunger gates were virtualized with respect to one another, which is not a necessary requirement for operation. In this regime, the virtual plunger and barrier amplitudes were set to 130 mV and 150 mV, respectively. 

\subsubsection{Single-qubit RB gateset}

Randomized benchmarking of the conveyor EDSR gate in Fig.~\ref{fig:fig2}f was performed using the X/Z compilation (Table~II in~\cite{xue_benchmarking_2019}). The baseband control of spin qubits performed in this work natively uses only positive X and Z rotations~\cite{wang_operating_2024, unseld_baseband_2025}. Given the Larmor frequencies at the conveyor EDSR condition (Fig.~\ref{fig:fig2}d), the $Z_{90}$ gate duration falls below 2~ns, which proved too short given the AWG bandwidth, resulting in a reduced Clifford gate fidelity. The $Z_{90}$ was therefore replaced by the equivalent $Z_{450}$ gate, obtained by waiting one additional full Larmor precession period. Table~\ref{tab:clifford_xz} lists the resulting Clifford compilation in the $\{X_{90}, Z_{180}, Z_{270}, Z_{450}\}$ gate set, where each Clifford contains exactly two $X_{90}$ gates. The $X_{90}$ gate error is therefore bounded from above by half the Clifford error.

\begin{table}[h]
\centering
\caption{Compilation of the single-qubit Clifford group with positive X and Z rotations, where $Z_{90}$ is replaced by the equivalent $Z_{450}$. Negative gates are decomposed into positive rotations, for example $-X_{90} \to Z_{180}, X_{90}, Z_{180}$ and $-Z_{90} \to Z_{270}$. The gateset is therefore \{$X_{90}$,\, $Z_{180}$,\, $Z_{270}$,\,$Z_{450}$\}, with every Clifford containing two $X_{90}$ gates.}

\label{tab:clifford_xz}
\renewcommand{\arraystretch}{1.3}
\begin{tabular}{l@{\hspace{1.5cm}}l}
\hline\hline
Class & $X^+/Z^+$ generation \\
\hline
Pauli
  & $X_{90}$, $Z_{180}$, $X_{90}$, $Z_{180}$ \\
  & $X_{90}$, $X_{90}$ \\
  & $Z_{270}$, $X_{90}$, $X_{90}$, $Z_{450}$ \\
  & $X_{90}$, $Z_{180}$, $X_{90}$ \\[4pt]
$2\pi/3$
  & $X_{90}$, $Z_{270}$, $X_{90}$, $Z_{450}$ \\
  & $X_{90}$, $Z_{450}$, $X_{90}$, $Z_{270}$ \\
  & $Z_{180}$, $X_{90}$, $Z_{450}$, $X_{90}$, $Z_{450}$ \\
  & $Z_{180}$, $X_{90}$, $Z_{270}$, $X_{90}$, $Z_{270}$ \\
  & $Z_{270}$, $X_{90}$, $Z_{450}$, $X_{90}$ \\
  & $Z_{270}$, $X_{90}$, $Z_{270}$, $X_{90}$, $Z_{180}$ \\
  & $Z_{450}$, $X_{90}$, $Z_{270}$, $X_{90}$ \\
  & $Z_{450}$, $X_{90}$, $Z_{450}$, $X_{90}$, $Z_{180}$ \\[4pt]
$\pi/2$
  & $Z_{270}$, $X_{90}$, $Z_{450}$, $X_{90}$, $Z_{270}$ \\
  & $Z_{270}$, $X_{90}$, $Z_{270}$, $X_{90}$, $Z_{270}$ \\
  & $X_{90}$, $Z_{270}$, $X_{90}$, $Z_{180}$ \\
  & $X_{90}$, $Z_{450}$, $X_{90}$, $Z_{180}$ \\
  & $Z_{180}$, $X_{90}$, $Z_{180}$, $X_{90}$, $Z_{450}$ \\
  & $Z_{180}$, $X_{90}$, $Z_{180}$, $X_{90}$, $Z_{270}$ \\[4pt]
Hadamard
  & $X_{90}$, $Z_{270}$, $X_{90}$ \\
  & $X_{90}$, $Z_{450}$, $X_{90}$ \\
  & $Z_{270}$, $X_{90}$, $Z_{450}$, $X_{90}$, $Z_{450}$ \\
  & $Z_{270}$, $X_{90}$, $Z_{270}$, $X_{90}$, $Z_{450}$ \\
  & $X_{90}$, $X_{90}$, $Z_{450}$ \\
  & $X_{90}$, $X_{90}$, $Z_{270}$ \\
\hline\hline
\end{tabular}
\end{table}

\section*{\label{sec:Supp}Supplementary information}

\setcounter{section}{0}
\setcounter{subsection}{0}
\setcounter{figure}{0}
\setcounter{table}{0}
\setcounter{equation}{0}

\renewcommand{\thesubsection}{\Alph{subsection}}

\renewcommand{\thefigure}{S\arabic{figure}}
\renewcommand{\thetable}{S\arabic{table}}
\renewcommand{\theequation}{S\arabic{equation}}

\renewcommand{\figurename}{FIG.}

\subsection{\label{supp:subsec:Rabi_decay}Rabi decay time during conveyor EDSR}

\begin{figure}[H]
\centering
\includegraphics[width=0.44\textwidth]{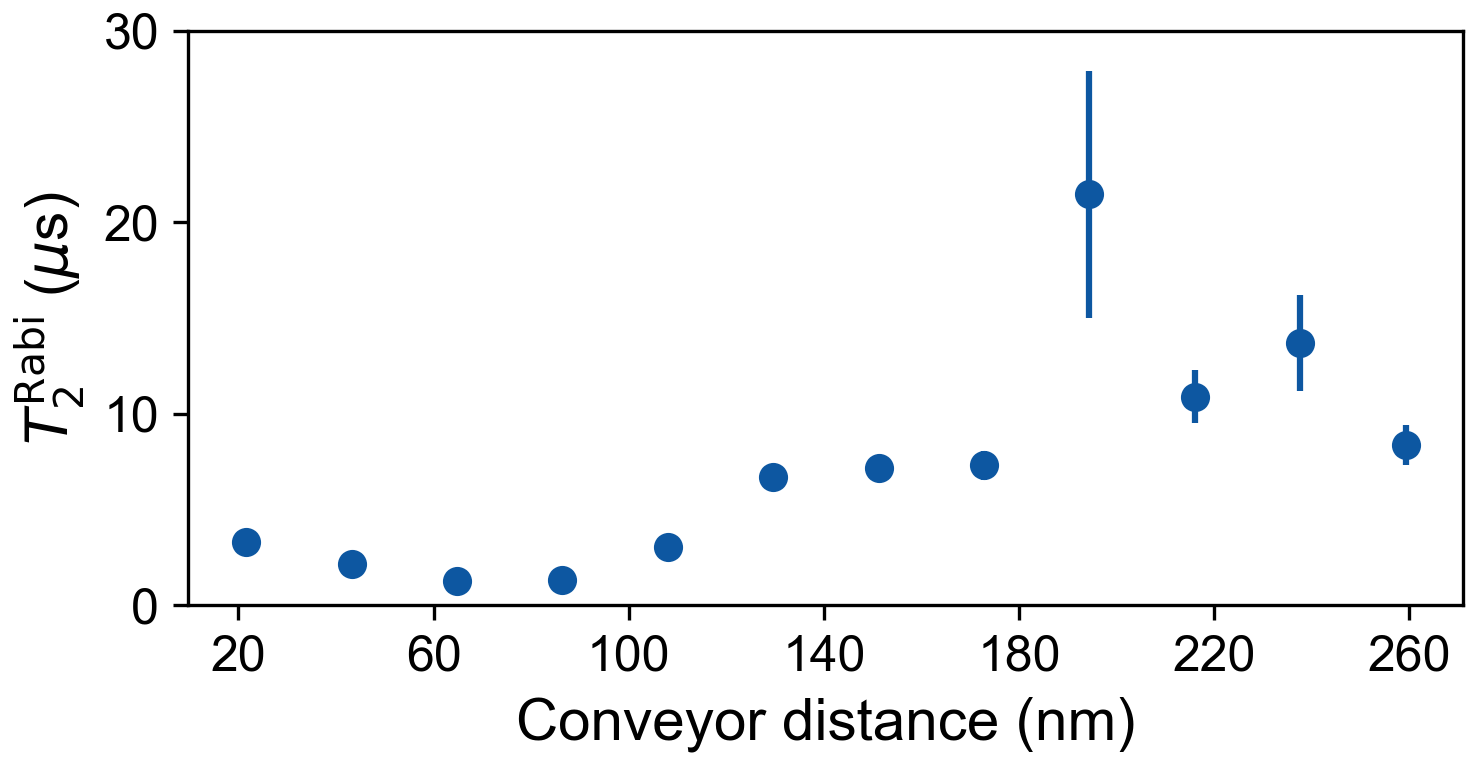}
\caption{\label{supp:fig:Rabi_decay} \textbf{Conveyor EDSR Rabi decay time.} The decay time of Q2 during conveyor EDSR. As in main Fig.~\ref{fig:fig2}c, the electron starts under gate P5 and is shuttled towards gate P2.
}
\end{figure}

Increasing the back-and-forth shuttle distance boosts the Rabi frequency, as demonstrated in main Fig.~\ref{fig:fig2}c. Fig.~\ref{supp:fig:Rabi_decay} further shows that the Rabi decay time $T_2^{Rabi}$ increases simultaneously with shuttle distance. Notably, $T_2^{Rabi}$ improves most significantly for shuttle distances exceeding $\sim$100 nm.  This length scale corresponds to a typical characteristic spatial correlation length of charge noise~\cite{rojas-arias_scaling_2026}, beyond which the motional averaging becomes effective.

\subsection{\label{supp:subsec:T2*}Coherence in static conveyor}

\begin{figure}[h]
\centering
\includegraphics[width=0.48\textwidth]{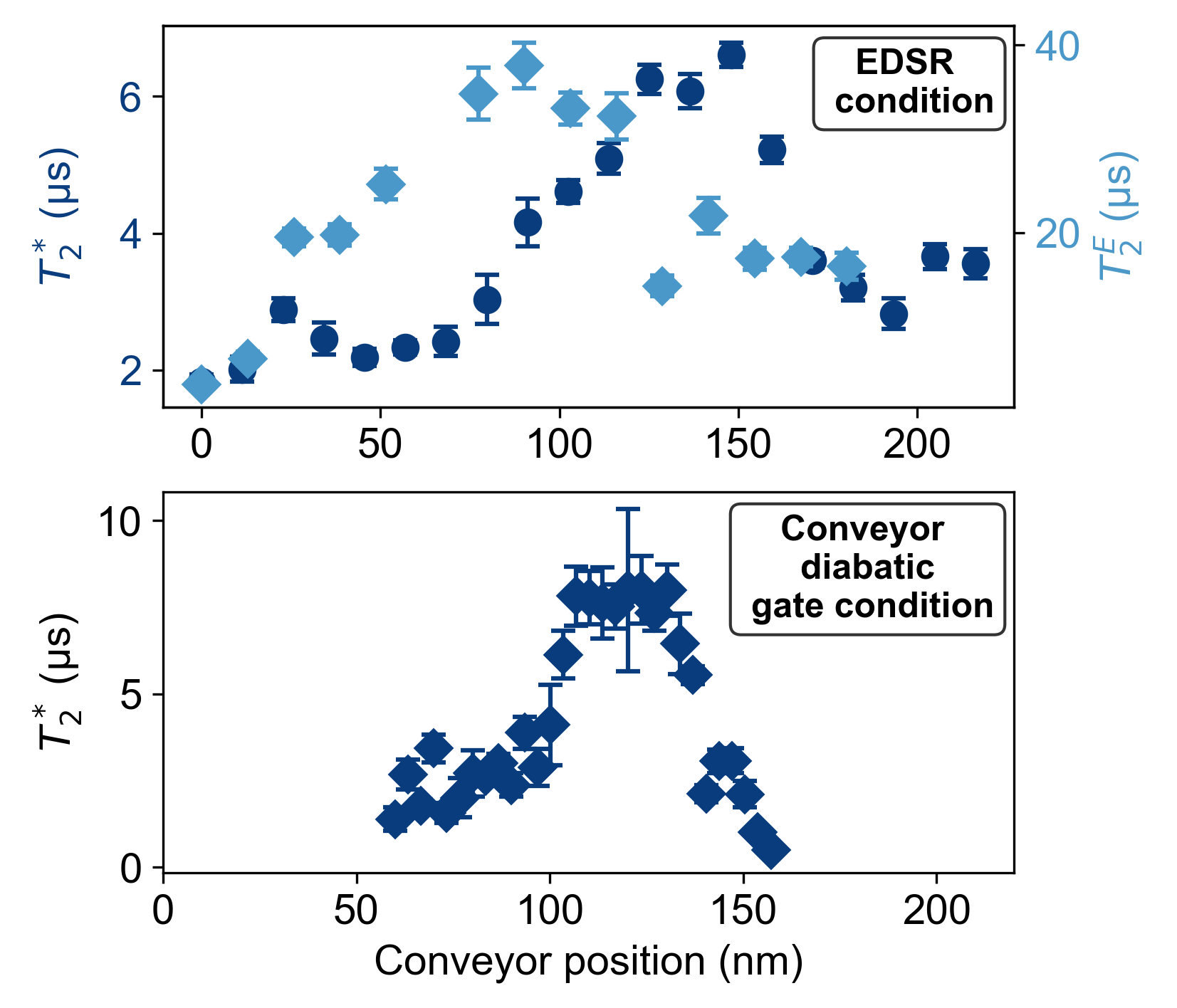}
\caption{\label{supp:fig:T2} \textbf{Coherence times for the conveyor EDSR (top) and diabatic (bottom) condition.} $T_2^{*}$ and $T_2^{E}$ of Q2 in the static conveyor potential.
}
\end{figure}

We extract the Ramsey $T_2^{*}$ and Hahn-echo $T_2^{E}$ decay times of Q2 idling in a static conveyor potential at varying positions along the array, with the zero position corresponding to 54 nm to the right of the center of gate P2. Fig.~\ref{supp:fig:T2} reveals substantial spatial variations in coherence times across the device. The measurements are performed under two distinct conditions, corresponding to the different magnetic fields used for the conveyor EDSR data of Fig.~\ref{fig:fig2} and the conveyor diabatic gate data of Figs.~\ref{fig:fig3} and~\ref{fig:fig4}.

Notably, under the EDSR condition, the $T_2^{*}$ and Hahn-echo $T_2^{E}$ do not peak at the same position, suggesting that the dynamical decoupling is not uniformly effective along the array. This could be explained by different noise power spectral densities versus position. For the diabatic condition, the exchange coupling with Q5 is activated at around 145 nm, significantly lowering the $T_2^{*}$.

\subsection{\label{supp:subsec:demagnetization}Micromagnet simulation and demagnetization}


\begin{figure}[h]
\centering
\includegraphics[width=0.4\textwidth]{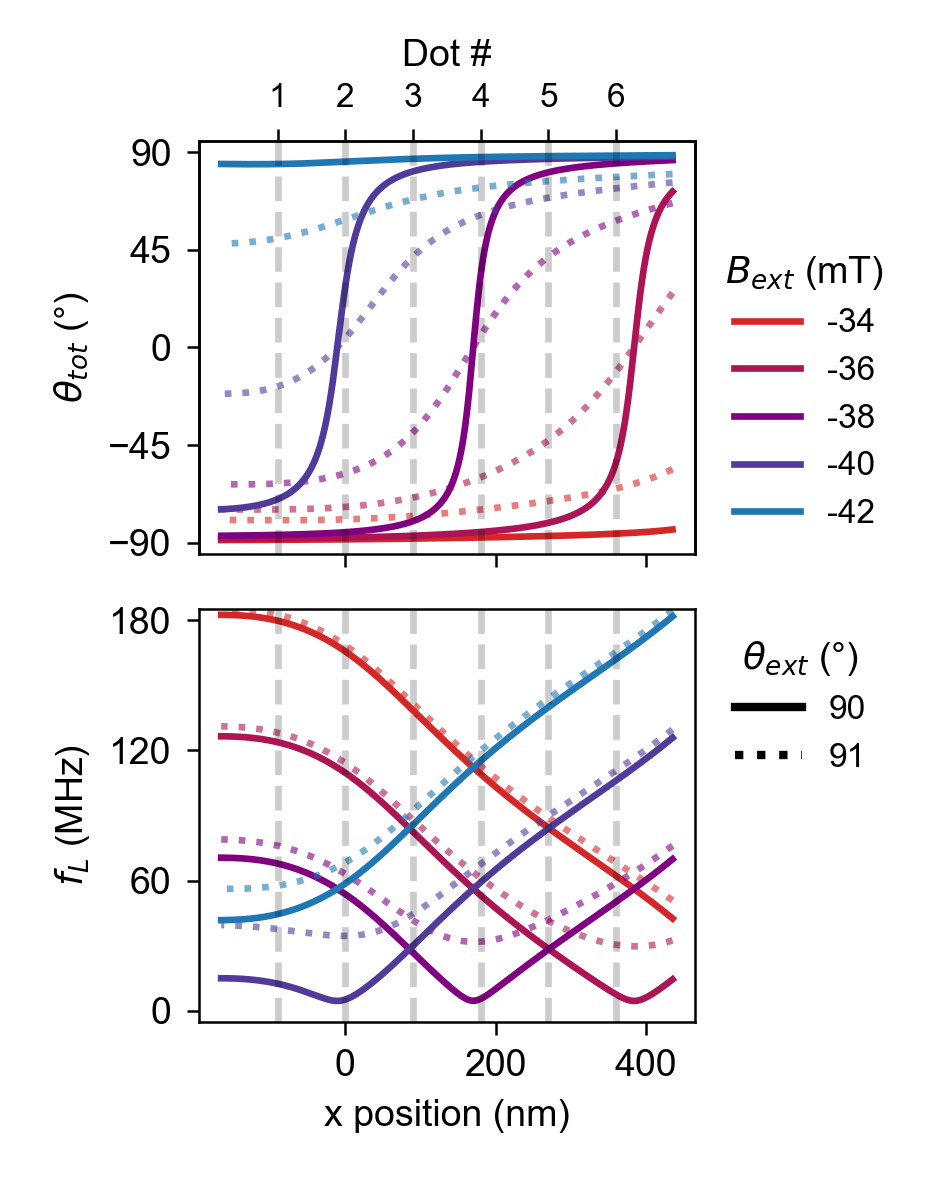}
\caption{\label{supp:fig:MMsim} \textbf{Simulation of the quantization-axis tilt and Larmor frequency along the linear array for diabatic conveyor shuttling.} \textbf{Top:} Polar angle $\theta_{tot}$ of the total magnetic field as a function of the x position, where $0\degree$ corresponds to the out-of-plane y direction and $\pm 90\degree$ corresponds to the $\pm$ z direction. \textbf{Bottom:} Larmor frequency $f_L$ along x, with quantum dot positions marked by vertical dashed lines. Colors indicate different values of $B_{ext}$. Solid lines correspond to $B_{ext}$ applied along the ideal z direction, which is the in-plane direction connecting the micromagnet slabs. The dotted lines indicate the effect of a 1 degree out-of-plane rotation in the z-y plane.}
\end{figure}

\begin{figure}[h]
\centering
\includegraphics[width=0.38\textwidth]{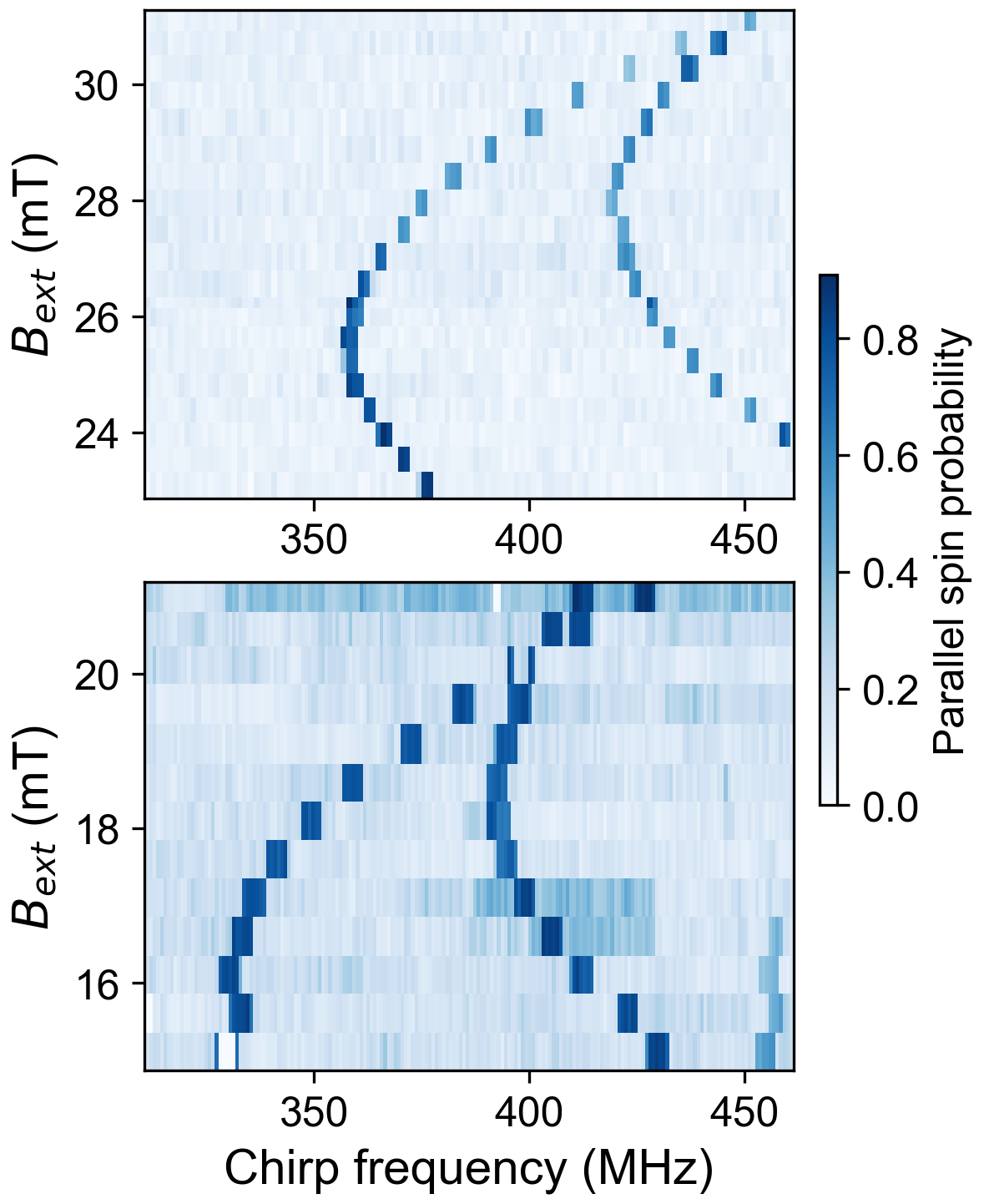}
\caption{\label{supp:fig:MW_chirp} \textbf{Demagnetization of the micromagnet.} Adiabatic inversion of stationary qubit 1 and 2 using a chirped microwave burst at different external magnetic field values $B_{ext}$. \textbf{Top:} After magnetizing the micromagnet at -2 T, the field is set to 34.5 mT and swept down. The chirped microwave pulse has a bandwidth of 3 MHz and duration of \SI{100}{\micro \second}. \textbf{Bottom:} Here, the micromagnet was magnetized at -1.8 T, and the field is then set to 36 mT, inducing more demagnetization, before being swept down.
}
\end{figure}

We simulate the tilt in the quantization axis and Larmor frequency as a function of the position along the array in Fig.~\ref{supp:fig:MMsim} using the Python package magpylib~\cite{ortner_magpylib_2020}. The remanence of the micromagnet is thereby used as a free scaling factor for the Larmor frequencies, whereby we assume in the simulation that the (remnant) magnetization is uniform across the micromagnet. The quantization axis and Larmor frequency for $B_{ext}$ close to -40 mT are qualitatively consistent with the experimental data in Fig.~\ref{fig:fig3}b and Fig.~\ref{fig:fig3}c, respectively. However, the simulation does not take into account the demagnetization of the micromagnet, making quantitative agreement hard. Importantly, the simulation reveals the sensitivity to field misalignment. When the external magnetic field is rotated from the z direction to 1 degree in the out-of-plane y direction, the maximum spatial gradient of the quantization angle decreases. At $B_{ext} =$ -38 mT, the field misalignment causes a reduction from 3.85 $\degree$/nm to 0.56 $\degree$/nm. Additionally, the Larmor frequency minimum is shifted away from zero. This provides a likely explanation for the data in Fig.~\ref{supp:fig:MW_chirp}. We note that similar results are obtained for an external field misalignment in the x direction.

After magnetizing the micromagnet by taking the externally applied magnetic field to -2 Tesla, we bring the external magnetic field to the few mT regime in the direction opposite to the initial magnetizing field. Using adiabatic spin inversion with chirped microwave control, we track the evolution of the qubit resonance frequencies of the stationary spins in QD 1 and QD 2 as a function of the external magnetic field $B_{ext}$, shown in Fig.~\ref{supp:fig:MW_chirp}, expecting the qubit frequency to pass through zero. Unexpectedly, the qubit frequencies exhibit minima in the range of a few hundreds of MHz. This indicates that the micromagnet stray magnetic field at the location of the qubits is not fully aligned with the direction defined by the external field orientation. 
Instead, a substantial transverse field component remains, likely arising from a combination of micromagnet stray fields and a small misalignment of the external magnetic field. To reach lower qubit frequencies necessary for baseband conveyor-controlled operations, we intentionally dynamically demagnetize the micromagnet. This is achieved by repeatedly sweeping the external field back and forth while gradually increasing the maximum field applied in the demagnetizing direction. As a result, the qubit frequency minima shift to lower values and occur at smaller $B_{ext}$, as illustrated in the bottom panel of Fig.~\ref{supp:fig:MW_chirp}. Rather than following a single asymmetric hysteresis cycle defined by the major loop and a first-order reversal curve, the repeated field sweeps progressively demagnetize the micromagnet. This produces non-closing minor hysteresis loops~\cite{moree2023review}, allowing us to gradually lower the magnetization and thus the total magnetic field experienced by the qubits.


\subsection{\label{supp:subsec:EDSR_Q5}Q5 conveyor EDSR}

\begin{figure}[h]
\centering
\includegraphics[width=0.48\textwidth]{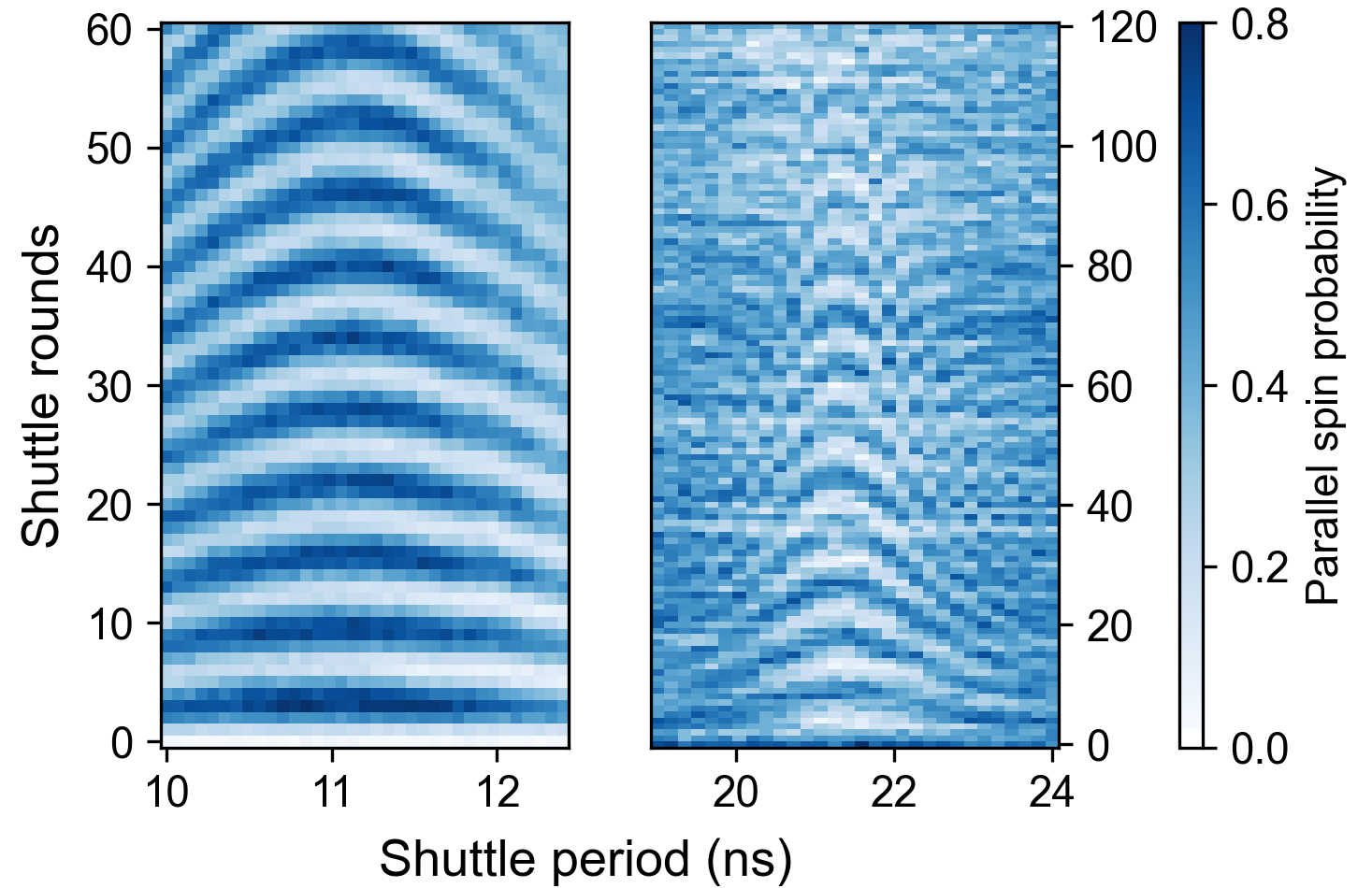}
\caption{\label{supp:fig:Supp_Chevron_Q5} \textbf{Conveyor EDSR of qubit 5.} \textbf{Left:} shuttling back-and-forth 97.2 nm at the local mean Larmor frequency drives Q5. This condition was used for Fig.~\ref{fig:fig4} of the main text. \textbf{Right:} Subharmonic resonance created by shuttling back-and-forth 86.4 nm at half of the local mean Larmor frequency.
}
\end{figure}

In Fig.~\ref{fig:fig4} of the main text, we load and control a spin in quantum dot 5. Because the region of strong quantization-axis tilt is located very close to static quantum dot 2, we drive Q5 with conveyor EDSR rather than diabatic shuttling. Figure~\ref{supp:fig:Supp_Chevron_Q5} shows the resonant chevron pattern obtained from back-and-forth shuttling of qubit 5. Interestingly, the qubit can also be resonantly driven by shuttling at half the Larmor frequency, although with lower Rabi frequencies. These subharmonic resonances were also observed for Q2 and at any low magnetic field condition. They are explained within the theoretical framework of~\cite{krzywda_coherence_2026}. Additional data on these subharmonics can be found in Supplementary Information~\ref{supp:subsec:subharmonics}.

\subsection{\label{supp:subsec:mod_CV}Conveyor EDSR with different motion profiles}

\begin{figure}[H]
\centering
\includegraphics[width=0.4\textwidth]{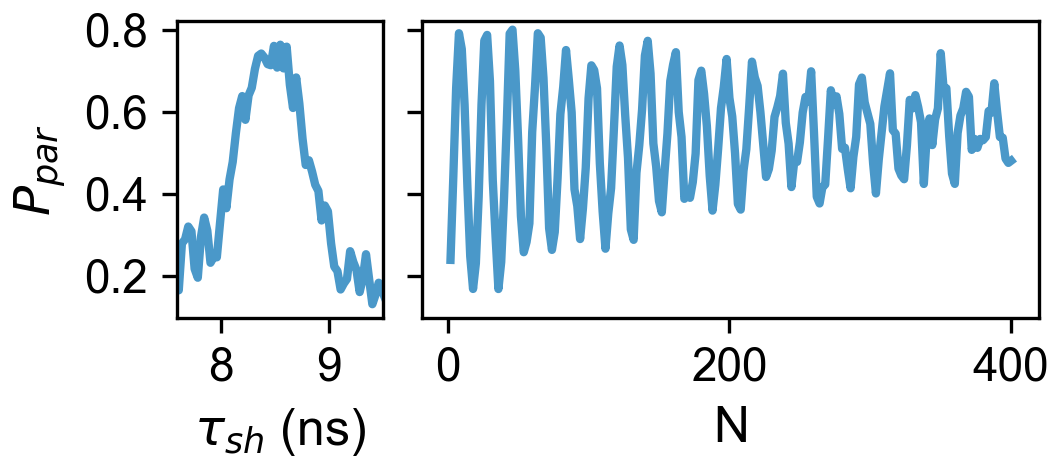}
\caption{\label{supp:fig:modulated_EDSR} \textbf{Conveyor EDSR of Q2 with frequency-modulated conveyor voltages for a sinusoidal motion profile (see Eq.~\ref{eq:smooth_profile}).} \textbf{Left:} Conveyor EDSR resonance showing the parallel spin probability of Q1-Q2 versus the shuttle period $\tau_{sh}$ after a fixed 16 shuttle rounds of 216 nm. \textbf{Right:} Resonant Rabi driving by varying the number of shuttle rounds $N$.  
}
\end{figure}

For simplicity, we normally operate conveyor EDSR by shuttling at constant speed with an immediate reversal of direction. This creates a triangular position profile in time. One could try to limit the Fourier components of the drive by producing a smooth, sinusoidal motion profile. Figure~\ref{supp:fig:modulated_EDSR} shows conveyor EDSR measurements, where this is accomplished by modulating the speed of the conveyor potential. Here, we operated a single tone four-phase conveyor instead of the two-tone conveyor~\cite{de_smet_high-fidelity_2025} used throughout this work. We encountered practical issues with the two-tone speed-modulated implementation, as now the gates receive many frequency components that excite the RF readout resonators. To achieve a sinusoidal speed evolution, the original sinusoidal gate voltages are given an argument that is itself frequency-modulated. The voltage applied to gate $n$ is 
\begin{equation}
\label{eq:smooth_profile}
V_n(t) = V^{DC}_n + A_n\sin \left(\frac{\pi d}{\lambda} \left[ 1-\cos \left(\frac{2\pi t}{\tau_{sh}} \right)  \right]  - \phi_n \right). 
\end{equation}

Here, $d$ is the one-way shuttle distance, $\lambda$ is the conveyor wavelength, and $\tau_{sh}$ is the back-and-forth shuttle period.\\

\begin{figure}[h]
\centering
\includegraphics[width=0.3\textwidth]{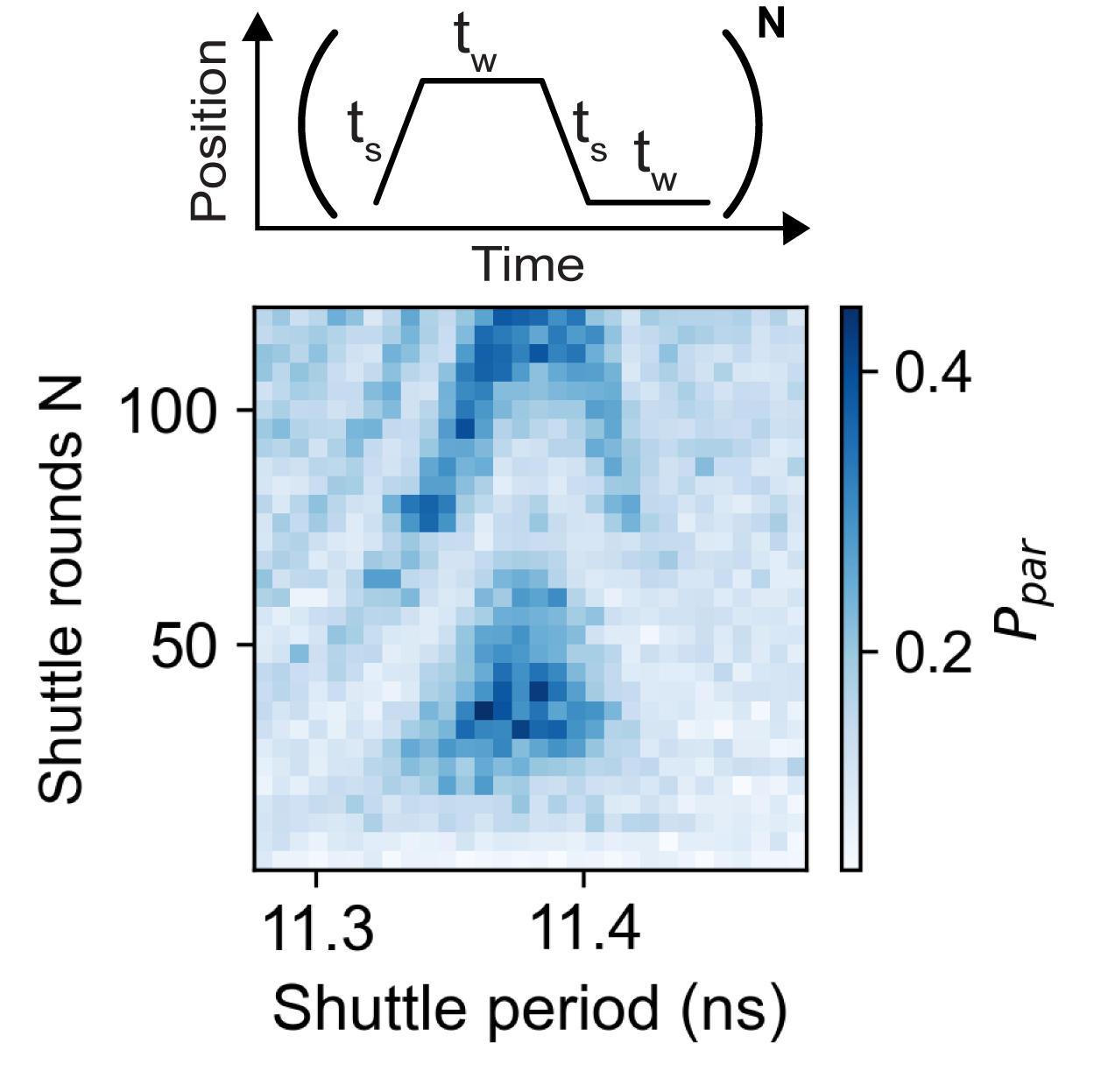}
\caption{\label{supp:fig:block_chevron} \textbf{Conveyor EDSR chevron with block motion profile.} \textbf{Top:} Schematic of $n$ shuttle rounds with a block-shaped qubit position as a function of time. The qubit starts at 216 nm and ends at 0 nm. As in the main text, the zero position is defined as the initial conveyor position, located 54 nm to the right of the center of gate P2. The one-way shuttle time $t_s$ is fixed at \SI{5}{\nano \second}, while the wait time per position $t_w$ is varied. \textbf{Bottom:} The resulting resonant conveyor EDSR chevron of Q2.}
\end{figure}

Next, we consider a block-like motion profile in which the electron is shuttled between two positions with a fixed one-way shuttle time $t_s$ and with variable wait time $t_w$ at each position. The total wait time is equally distributed between the two end positions. The resulting shuttle period is therefore $\tau_{sh} = 2 (t_s+t_w)$. Interestingly, this block motion profile also produces resonant conveyor EDSR chevron patterns, as shown in Fig.~\ref{supp:fig:block_chevron}.

\subsection{\label{supp:subsec:subharmonics}Subharmonic resonances of conveyor EDSR}

\begin{figure}[h]
\centering
\includegraphics[width=0.35\textwidth]{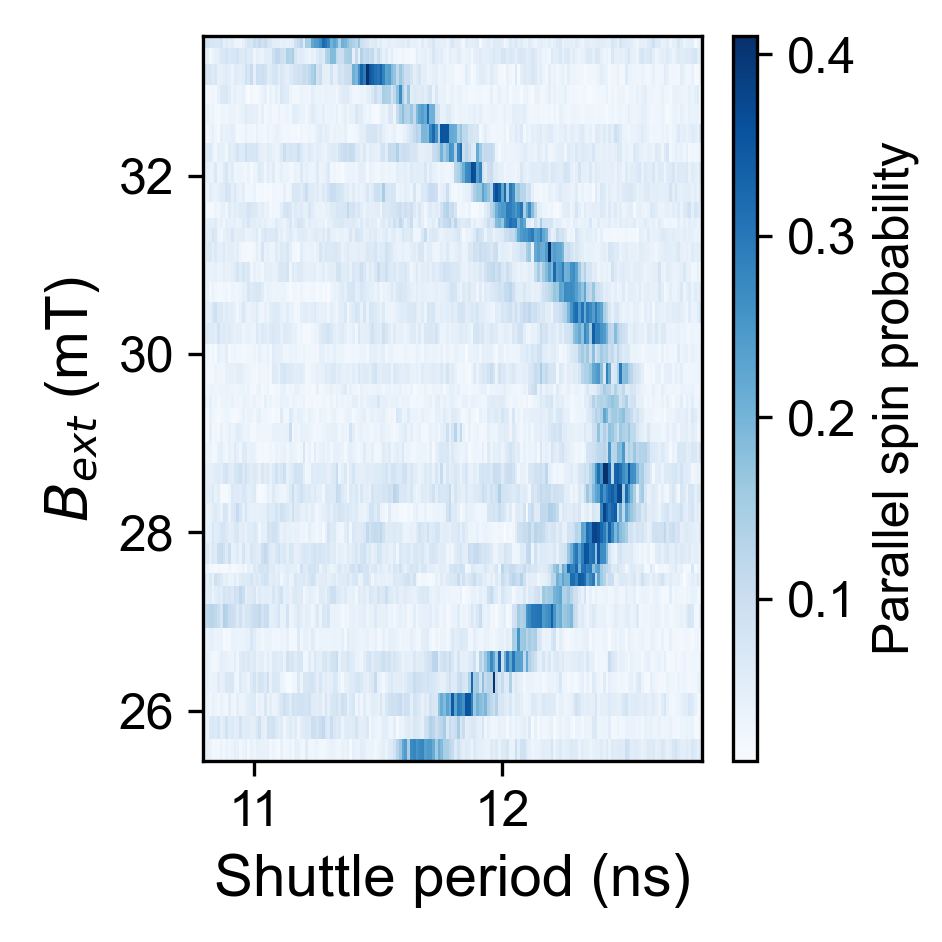}
\caption{\label{supp:fig:EDSR_shuttling_kink} \textbf{Resonant shuttling at a subharmonic of the mean Larmor frequency.} Conveyor EDSR resonance of Q2 obtained after 10 shuttle rounds with a triangular motional profile, showing a shift with the external magnetic field $B_{ext}$. The micromagnet condition is identical to that used in the top panel of Fig~\ref{supp:fig:MW_chirp}.}
\end{figure}

In Supplementary Information~\ref{supp:subsec:demagnetization}, we reduce the external magnetic field to lower the qubit frequencies. Without repeated demagnetization of the micromagnet, the minimum achievable qubit frequencies remain in the several hundred MHz, which is beyond the accessible range of our back-and-forth shuttling frequencies. Nonetheless, in Figure~\ref{supp:fig:EDSR_shuttling_kink}, we observe resonant conveyor EDSR control of Q2 at a significantly lower shuttling frequency of around 80 MHz (12.5 ns shuttle period) within the same $B_{ext}$ range as in Fig.~\ref{supp:fig:MW_chirp}. This discrepancy indicates that the observed resonance does not correspond to the fundamental Larmor frequency. Instead, we attribute it to a higher-order subharmonic, which can be driven by the strong longitudinal field modulation in the Floquet frame~\cite{krzywda_coherence_2026}. As expected, the Rabi frequency is significantly lower for higher-order subharmonic resonances than for the fundamental resonance.

\begin{figure}[h]
\centering
\includegraphics[width=0.48\textwidth]{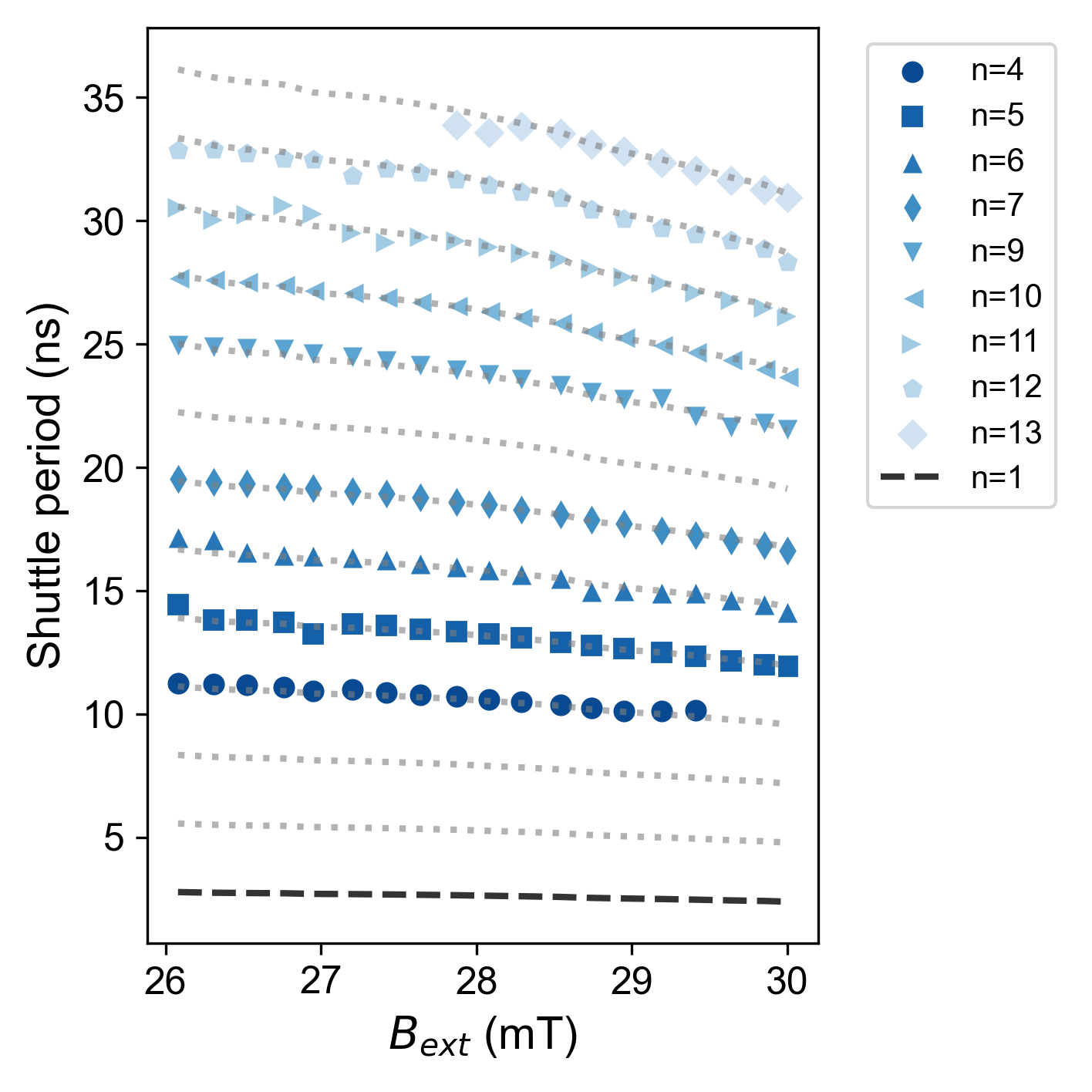}
\caption{\label{supp:fig:subharmonics} \textbf{Conveyor EDSR subharmonics when shuttling with a block-like motion profile.} The one-way shuttle time $t_s$ is fixed at \SI{5}{\nano \second}, while the shuttle period is varied by sweeping the waiting time at both positions simultaneously. Markers indicate experimentally obtained resonant periods. The mean Larmor frequency $f_0$ (black dashed line) is extracted from the average spacing between subharmonic resonances, which then enables assignment of the integer indices $n$. Gray dotted lines indicate the expected resonance positions of the subharmonics.}
\end{figure}

These subharmonic resonances are also observed when shuttling with a block-like motion profile (see Supplementary Information~\ref{supp:subsec:mod_CV}). A subharmonic is driven when the shuttle period satisfies $\tau_{sh} = n f_0^{-1}$, where $f_0^{-1}$ is the period corresponding to the mean Larmor frequency $f_0$ along the shuttling path, and $n$ is an integer. In Figure~\ref{supp:fig:subharmonics}, we extract the resonant shuttle periods over a range of external magnetic field values. Subharmonics corresponding to $n = 4$ to $n = 13$ (excluding $n = 8$) are clearly identified, while in some datasets subharmonics up to $n = 21$ can be resolved. The fundamental resonance $n = 1$ at $f_0$ is not observed due to limitations in the achievable shuttle speed for this distance. Resonances $n = 2$ and $n = 3$ are also absent, as the minimum achievable shuttle period for this motion profile was set by the fixed total shuttle time $2 t_s$ of \SI{10}{\nano \second}. Nevertheless, the mean Larmor frequency $f_0$ can be inferred from the average spacing between successive subharmonic resonances at each value of $B_{ext}$, yielding values between 344 and 424 MHz. For this dataset, the micromagnet was further demagnetized compared to the condition used in Fig.~\ref{supp:fig:EDSR_shuttling_kink}.

\subsection{\label{supp:subsec:IRB}Interleaved RB for conveyor EDSR}

\begin{figure}[h]
\centering
\includegraphics[width=0.44\textwidth]{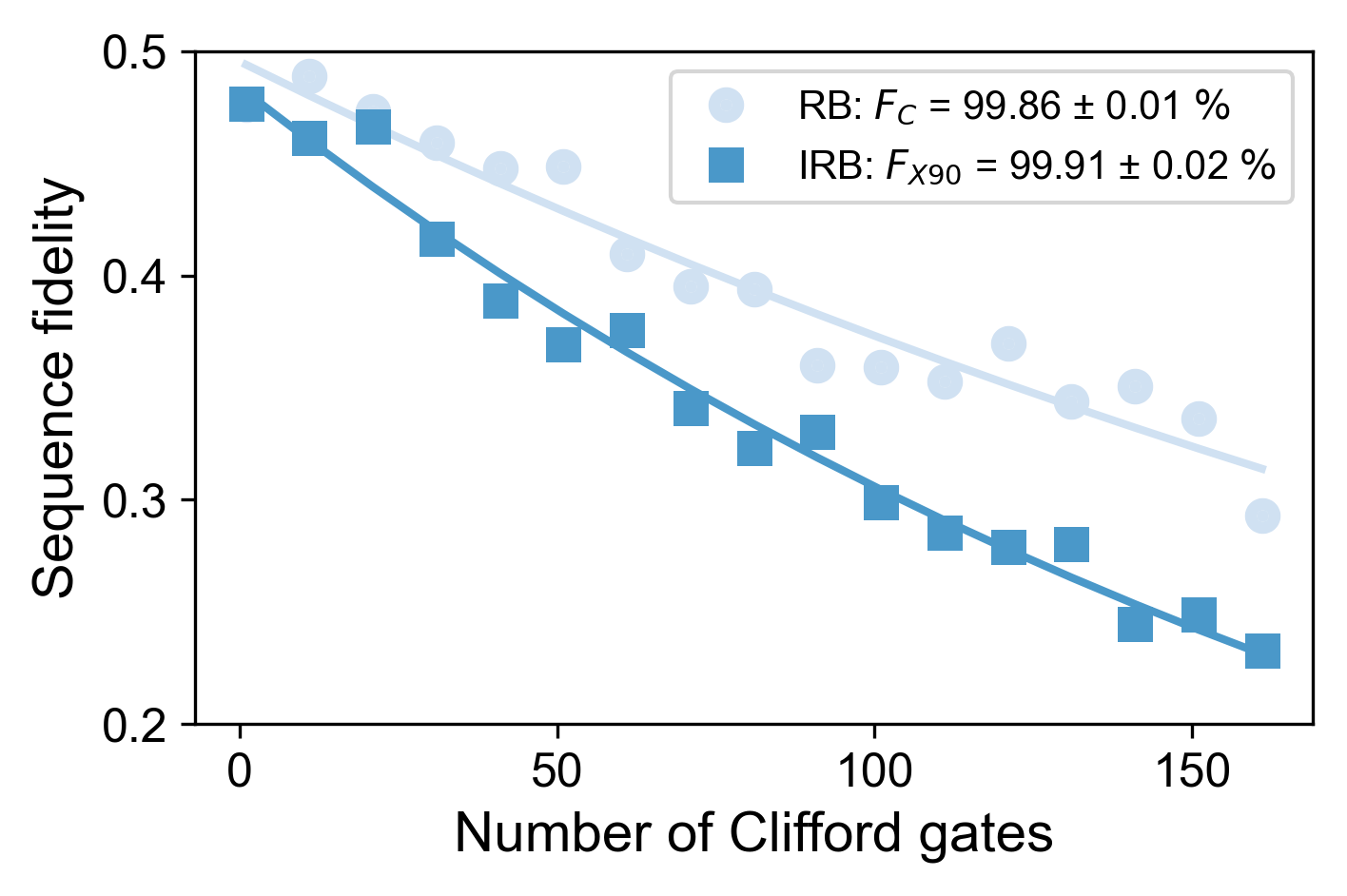}
\caption{\label{supp:fig:IRB} \textbf{Conveyor EDSR interleaved RB.} The sequence fidelity is obtained from the difference between RB measurements with input states $\ket{\uparrow}$ and $\ket{\downarrow}$, each averaged over 40 randomizations. From the $X_{90}$ interleaved sequence we extract a gate fidelity $F_{X90} = 99.91 \pm 0.02\%$. The uncertainty represents one standard deviation extracted from the exponential fit.}
\end{figure}

In Fig.~\ref{fig:fig2}f of the main text, we benchmark conveyor EDSR single-qubit operations using randomized benchmarking (RB) and report a Clifford fidelity of $99.84 \pm 0.01\%$. In previous work on shuttling-based gates~\cite{wang_operating_2024, unseld_baseband_2025}, not all gates in the gateset yielded similar fidelities, making it difficult to reliably extract an average gate fidelity directly from the Clifford fidelity. In particular, the $X_{90}$ gate relies on shuttling and is expected to be the dominant contributor to the Clifford error. Figure~\ref{supp:fig:IRB} shows interleaved randomized benchmarking (IRB) of the $X_{90}$ gate. The reference RB measurement yields a Clifford fidelity of $99.86 \pm 0.01\%$, in good agreement with the value reported in the main text. Assuming that the Clifford error is dominated by the two $X_{90}$ gates per Clifford, we estimate a gate fidelity of $F_{X90} = 99.93\%$. This estimate is consistent with the IRB result of $99.91 \pm 0.02\%$ within the uncertainty. We therefore conclude that the three rounds of back-and-forth shuttling required for the $X_{90}$ gate constitute the dominant contribution to the Clifford error, rather than the short idle intervals used to implement Z rotations.

The arbitrary waveform generator used in these experiments has a sample rate of 1 GSa/s, while the pulse compiler uses interpolation to enhance resolution to 0.1 ns. Since Z rotations are implemented through idling times, the Z gates in the RB sequence have a uniformly distributed timing error in the $\pm$ 50 ps range. For the conveyor EDSR case in main Fig.~\ref{fig:fig2}e, where the Larmor period is 7.7 ns, this leads to a maximum phase error of 0.0408 rad and an rms phase error of $\sqrt{\langle \delta\phi^2 \rangle} = \delta\phi_{\mathrm{max}}/\sqrt{3}$. A simple estimate of the resulting infidelity $1-F \approx \langle\delta\phi^2\rangle/6 = 9\times10^{-5}$ shows that these phase errors are an order of magnitude smaller than the $X_{90}$ shuttling gate error, supporting the conclusion from the interleaved randomized benchmarking measurements.

\subsection{\label{supp:subsec:read_wait}Impact of wait time before readout on RB fidelity}

\begin{figure}[h]
\centering
\includegraphics[width=0.39\textwidth]{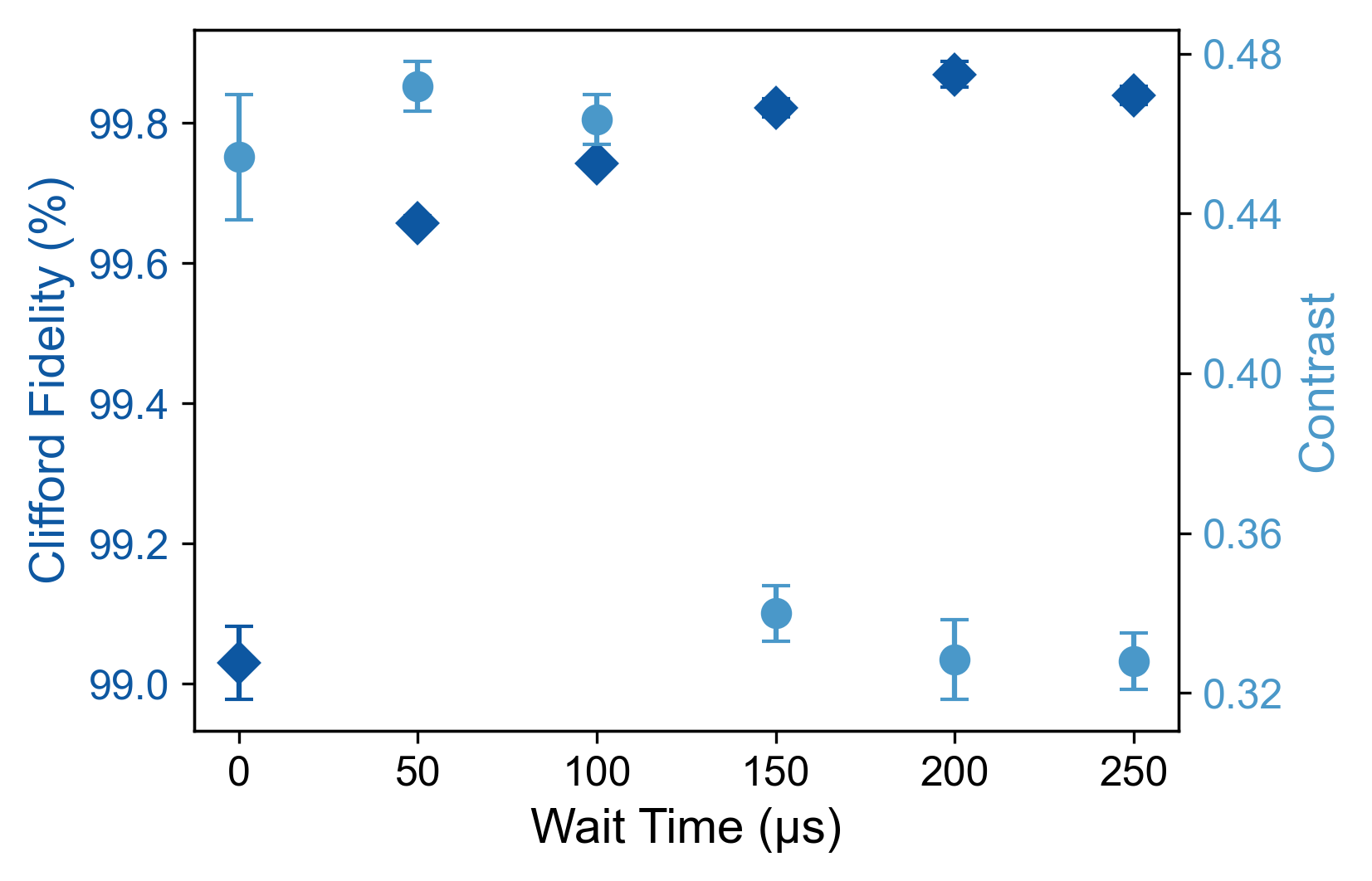}
\caption{\label{supp:fig:readwait} \textbf{Conveyor EDSR RB with varying wait time before readout.} After the RB sequence, a wait time in the (3,1) charge state is added before pulsing to the readout voltage configuration. The Clifford fidelity (diamond markers) improves with increased wait time, while the initial RB contrast between $\ket{\downarrow}$ and $\ket{\uparrow}$ (round markers) diminishes.}
\end{figure}

We consistently observe that, for long RB sequences, adding an extra wait time in the (3,1) charge configuration before readout improves the characteristic decay. Figure~\ref{supp:fig:readwait} shows that the extracted Clifford error is reduced by an order of magnitude when waiting \SI{200}{\micro \second} compared to adding no wait time. We speculate that the gate voltage pulses for back-and-forth shuttling contain many frequency components due to the instantaneous reversal of the electron motion, which can be close to the frequency of the RF readout resonators. Repeated shuttling for long RB sequences could therefore excite the RF readout resonators, leading to a deteriorating readout fidelity with increasing number of Clifford gates. Including a wait time before readout can then help to depopulate the readout resonator. At the same time, waiting for extended times before readout leads to a diminishing initial RB contrast between the initialized states $\ket{\downarrow}$ and $\ket{\uparrow}$. This seems to be consistent with $T_1$ decay during the added wait time, though this relaxation time would be faster than expected for a detuning far from the (3,1)-(4,0) transition.

\subsection{\label{supp:subsec:ST_osc} Singlet-triplet oscillations during conveyor interaction}

\begin{figure}[h]
\centering
\includegraphics[width=0.36\textwidth]{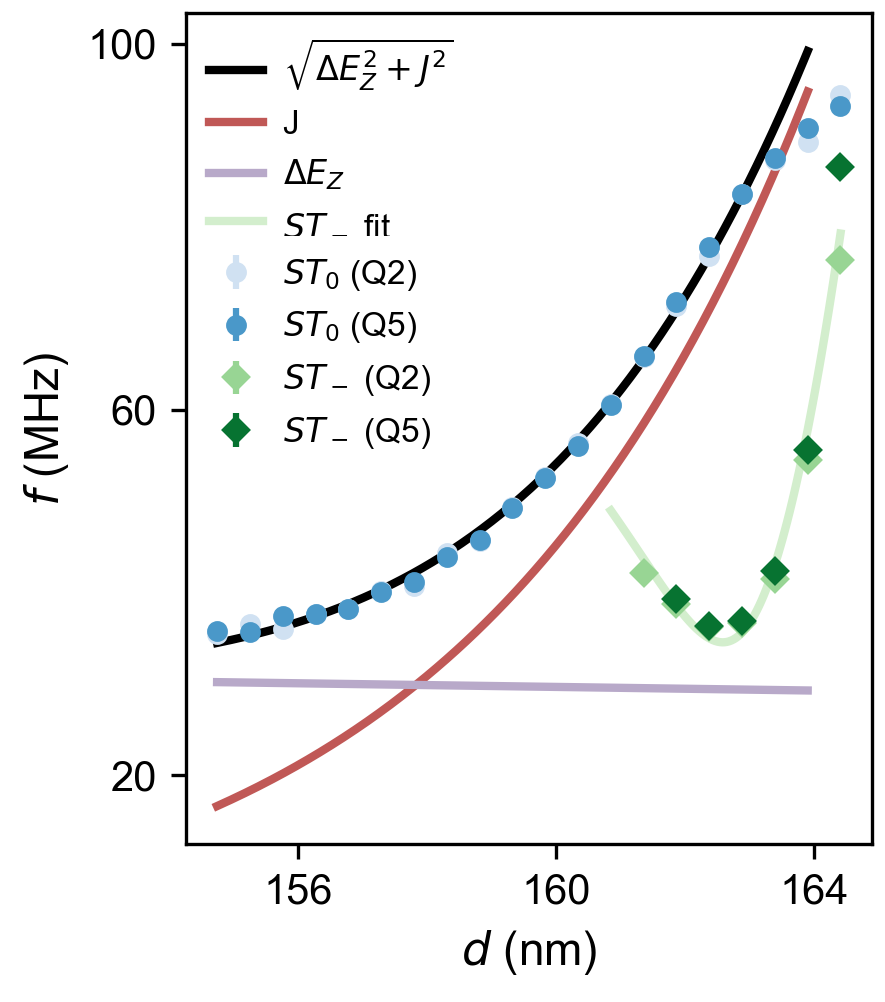}
\caption{\label{supp:fig:ST_osc} \textbf{Oscillation frequencies when Q2 is shuttled close to Q5.} The $ST_{0}$ markers correspond to the data of main Figure~\ref{fig:fig4}e, where the spins were initialized in the $\ket{\uparrow \downarrow}$ state. We fit the data with $\sqrt{\Delta {E_Z}^2 +J^2}$ (black line) as described in the text and visualize the components $J$ (red line) and $\Delta E_Z$ (purple line). When initializing in $\ket{\downarrow \downarrow}$, the oscillation frequency at the $ST_{-}$ anti-crossing is obtained. From the fit (green line) described in the text, we deduce the average Zeeman splitting $\overline{E}_Z$ and the artificial spin-orbit coupling $\Delta_{SO}$.}
\end{figure}

From the data of main Figure~\ref{fig:fig4}e, we extract the oscillation frequencies of Q2 and Q5, where the spins were initialized in the $\ket{\uparrow \downarrow}$ state. Both the exchange strength $J$ and the Zeeman splitting difference $\Delta E_Z$ contribute to these oscillations in the $ST_{0}$ subspace. Their frequencies can be fitted by $\sqrt{\Delta {E_Z(d)}^2 +J(d)^2}$. We approximate the Zeeman splitting difference to be linearly dependent on the distance between the qubits, which decreases as the shuttled distance $d$ of Q2 increases. The exchange strength is assumed to be exponentially increasing with $d$. In Figure~\ref{supp:fig:ST_osc} we plot the extracted oscillation frequencies and fit the $J$ and $\Delta E_Z$ contributions. \\

When the spins are initialized in the $\ket{\downarrow \downarrow}$ state, we reach the $ST_{-}$ anti-crossing when the exchange strength equals the average Zeeman splitting of the qubits. The $ST_{-}$ anti-crossing is fitted with $\sqrt{\Big(\overline{E}_Z-J\Big)^2 + \Delta_{SO}^2}$~\cite{farina_site-resolved_2025, zhang_universal_2025}, where we assume the average Zeeman splitting $\overline{E}_Z$ to be independent of $d$ in this small range. This yields a spin-orbit coupling $\Delta_{SO}$, artificially created by the micromagnet, of 34.6 MHz.

\subsection{\label{supp:subsec:diabatic_analysis}Analysis and simulations for diabatic shuttling gates}

\subsubsection{Model Hamiltonian}

The starting point is a minimal Hamiltonian for two electron-spin qubits subject to a spatially varying magnetic field vector, resulting from the vector sum of the externally applied magnetic field and the stray magnetic field from the micromagnet. The left qubit is shuttled towards the right qubit, which remains static. We set the $z$-axis parallel to the local magnetic field at the starting point of the left qubit. Restricting the dynamics to the two-spin subspace $\{\ket{\uparrow \uparrow}, \ket{\uparrow \downarrow}, \ket{\downarrow \uparrow}, \ket{\downarrow \downarrow}\}$, the Hamiltonian reads
\begin{equation}
	H(d) = \frac{J(d)}{4} \vec{\sigma}_L \cdot \vec{\sigma}_R + \frac{\mu_B g}{2} \vec{B}(x_L) \cdot \vec{\sigma}_L + \frac{\mu_B g}{2} \vec{B}(x_R) \cdot \vec{\sigma}_R
	\label{eq:Hamiltonian_2Q}
\end{equation}
where $\vec{\sigma}_i$ are the Pauli operators acting on qubit $i=L,R$, $\mu_B$ is the Bohr magneton, $g$ is the electron $g$-factor, and $J(d)$ is the distance-dependent exchange coupling. The inter-qubit distance is $d = x_R - x_L$, with $x_R$ fixed and $x_L$ varying over time during shuttling.

The central task is, therefore, to determine from the experiment two ingredients: the exchange coupling $J(d)$ and the total magnetic field at the two qubit positions, $\vec{B}_T(x_{i})$. The fitting strategy is hierarchical: we first calibrate the field sampled by the shuttled qubit, then use two-qubit datasets to infer the right-qubit field and the exchange profile, and finally use the calibrated model to predict the accessible two-qubit gate set. Whenever fitted interaction strengths are quoted in MHz below, the corresponding energy scale is implicitly divided by $h$. We begin with the field profile experienced by the moving qubit, as it can be isolated before addressing the full two-qubit problem.

\subsubsection{Single-qubit calibration} \label{sec:one_qubit_fitting}
The calibration begins in the simplest regime, where exchange is absent, and the magnetic-field profile can be probed directly. We therefore analyze a single electron-spin qubit shuttled along a one-dimensional channel in the presence of the total magnetic field $\vec{B}$. In this step, the left qubit is initialized in $\ket{\uparrow}$, shuttled forward, kept in a fixed position for a variable wait time, and then shuttled backward through the field gradient and measured in the $\{\ket{\uparrow},\ket{\downarrow}\}$ basis. The observed probability $P_{\downarrow}$ is recorded as a function of the wait time, conveyor distance, and shuttling speed.

We begin with a fixed shuttling velocity, $v = 16.2$~m/s, fast enough for the transfer to be predominantly diabatic with respect to the spin dynamics. The spin rotation accumulated during the waiting time between the forward and backward shuttles, therefore, probes the local magnetic field at the turning point.
For a fixed turning point, the single-qubit Hamiltonian is
\begin{equation}
	H_{1Q} = \frac{\mu_B g B}{2}(\cos(\theta) \sigma_z + \sin(\theta) \sigma_x),
	\label{eq:Hamiltonian_1Q}
\end{equation}
where $B$ is the total magnetic field at the turning point and $\theta$ is its polar angle with respect to the $z$ axis. We assume that the field lies in the $x$-$z$ plane, which is a good approximation for the micromagnet used in the experiments.
It is important to note that the direction labeled as $x$ might be arbitrary in the plane, as long as it changes sufficiently smoothly during motion.

For a waiting time $t_w$, the corresponding unitary evolution is
\begin{equation}
	U_{1Q}(t_w) = \exp\left[-i \pi f_L t_w (\cos(\theta) \sigma_z + \sin(\theta) \sigma_x)\right],
\end{equation} 
where $f_\mathrm{L}(x) = g \mu_B B(x)/h$ is the Larmor frequency.
The probability of measuring the spin in the down state after this evolution is
\begin{align}
	P^\mathrm{(ideal)}_{\downarrow}(t_w) &= \abs{\bra{\downarrow} U_{1Q}(t_w) \ket{\uparrow}}^2 \nonumber\\
	&= \frac{\sin^2(\theta)}{2} \left[1 - \cos(2\pi f_L t_w)\right].
	\label{eq:prob_down_1Q}
\end{align}
This equation immediately shows that if the total field is aligned with the $z$ axis ($\theta = 0, \pi$), then $P_{\downarrow}=0$, as expected. A finite transverse component ($\theta \neq 0$) induces precession around the total field. The oscillation amplitude is maximal when the field is perpendicular to the $z$ axis ($\theta = \pi/2, 3\pi / 2$), where the spin can rotate fully between $\ket{\uparrow}$ and $\ket{\downarrow}$.

To describe the measured signal, we include a phenomenological visibility factor $V$ and an offset $P_\mathrm{off}$ to account for SPAM imperfections. In addition, the finite shuttle duration can generate an effective phase accumulation, which we absorb into a phase offset $\phi$. The measured probability is then written as
\begin{align}
	P^\mathrm{(real)}_\downarrow(x, t_\mathrm{w}) = P_\mathrm{off} + \frac{V \sin^2\theta(x)}{2} \left[ 1 - \cos\left( 2 \pi f_{\mathrm{L}}(x) t_\mathrm{w} + \phi \right) \right].
	\label{eq:prob_down_1Q_real}
\end{align}
At this stage, $V$ and $\theta$ are not independently identifiable, so we fit only the product $A(x) \equiv V\sin^2\theta(x)$, which sets the oscillation amplitude. The oscillation frequency is determined by the field magnitude $B(x)$ through the Larmor relation.

\begin{figure}[t!]
	\centering
	\includegraphics[width=\columnwidth]{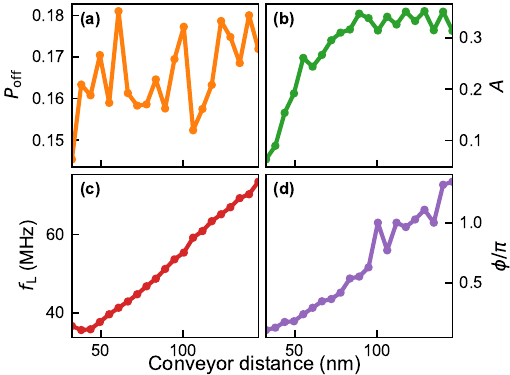}
	\caption{Single-qubit calibration data for two diabatic shuttles at fixed velocity.
	(a-d) Fitted offset, oscillation amplitude, Larmor frequency, and phase extracted from Fig.~\ref{fig:fig3}d~(left) of the main text as a function of the conveyor distance.
	These observables constitute the experimental input for reconstructing the local field sampled by the moving qubit.
	}
	\label{supp:fig:amplitude_fitting}
\end{figure}

The experimental data for the probability of measuring the spin in the $\ket{\downarrow}$ state after one diabatic shuttle round is shown in Fig.~\ref{fig:fig3}d~(left) of the main text, and the fitted parameters are reported in Fig.~\ref{supp:fig:amplitude_fitting}. The fitted offset $P_\mathrm{off} \approx 0.17$ remains nearly constant along the shuttling path. This observation supports treating the SPAM contribution, and therefore the effective visibility within this dataset, as approximately position independent. The amplitude $A$ increases with distance, indicating that the field tilts progressively away from the $z$ axis. The Larmor frequency is non-monotonic, with a minimum near $x\sim 40$~nm, consistent with micromagnet simulations. The phase offset $\phi$ also increases with distance, as expected if non-diabatic phase accumulation grows with the shuttle duration.

\begin{figure}[t!]
	\centering
	\includegraphics[width=\columnwidth]{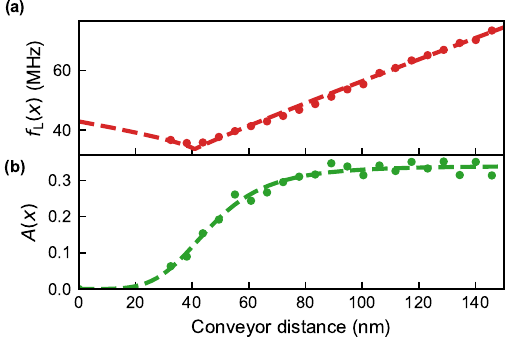}
	\caption{Phenomenological interpolation functions used in the time-dependent simulations.
	(a) Piecewise fit of the Larmor frequency $f_\mathrm{L}(x)$ to Eq.~(\ref{eq:extrapolation}).
	(b) Hill-type fit of the oscillation amplitude $A(x)$ to  Eq.~(\ref{eq:Hill-type}).
	The dashed curves are smooth interpolants constrained by the measured data points and by the limiting behavior discussed in the text.
	}
	\label{supp:fig:extrapolation}
\end{figure}

To separate the visibility from the polar angle, we first construct smooth interpolation functions for the Larmor frequency and the oscillation amplitude. For the Larmor frequency, we use the phenomenological piecewise form
\begin{equation}
	f_\mathrm{L}(x) = \begin{cases} 
		f_\mathrm{L}^{\min} + c_L (x_{\min} - x) ^{p_L} & \text{if } x < x_{\min} \\
		f_\mathrm{L}^{\min} + c_R (x - x_{\min}) ^{p_R}, & \text{if } x \geq x_{\min}
	\end{cases},
	\label{eq:extrapolation}
\end{equation}
where $f_\mathrm{L}^{\min}$ is the minimum Larmor frequency, $x_{\min}$ is the position at which the minimum is reached, and $c_L$, $p_L$, $c_R$, and $p_R$ determine the curvature on the two sides of the minimum. The best fit yields $f_\mathrm{L}^{\min} = 33.5$~MHz, $x_{\min} = 40.8$~nm, $c_L = 0.57$~MHz/nm$^{p_L}$, $p_L = 0.76$, $c_R = 0.48$~MHz/nm$^{p_R}$, and $p_R = 0.95$. As shown in Fig.~\ref{supp:fig:extrapolation}a, this form captures the observed non-monotonic dependence, including the minimum near $x\sim 40$~nm and the increase on both sides of that minimum. The form in Eq.~(\ref{eq:extrapolation}) is not intended as a microscopic model of the micromagnet; it is a compact interpolation that captures the data and provides stable extrapolation over the range required below.

For the oscillation amplitude, in addition to the values extracted from Fig.~\ref{supp:fig:amplitude_fitting}b, we impose $A(0) = 0$ because the field is parallel to the $z$ axis at $x=0$, and therefore $\theta(0) = 0$. The Hill-type function provides a good description of the data
\begin{equation}
	A(x) = A_{\max} \frac{x^n}{x^n + x_0^n},
	\label{eq:Hill-type}
\end{equation}
where $A_{\max}$ is the maximum amplitude, $x_0$ is the distance at which the amplitude reaches half of its maximum value, and $n$ controls the steepness of the crossover. The best fit gives $A_{\max} = 0.34$, $x_0 = 46$~nm, and $n = 4.47$, as shown in Fig.~\ref{supp:fig:extrapolation}b. Again, this form is not meant to be microscopic; it simply enforces the two experimentally motivated limits $A(0)=0$ and saturation at large $x$ while remaining smooth and monotonic.

To break the residual coupling between the visibility and the tilt angle, we use a protocol that is sensitive to the change in field orientation between $x=0$ and $x_{\max}$. The qubit is shuttled diabatically to $x_{\max}$, allowed to evolve for a time $t_1$, shuttled back to $x=0$, and allowed to evolve for a time $t_2$. This forward-backward sequence is repeated $n$ times before spin readout at $x=0$. Because the spin evolves around two non-collinear quantization axes, the final $P_\downarrow$ is highly sensitive to $\theta(x_{\max}) - \theta(0)$, making it possible to determine $V$ and $\theta(x)$ independently.

As a first approximation, if the shuttling speed is faster than the spin-precession timescale, the shuttle can be treated as perfectly diabatic, and the spin state is preserved during the motion. In that limit, the unitary time evolution for the multiple-shuttling protocol can be written as
\begin{subequations}
	\begin{equation}
		U_{t_1} = \exp\left[-i \pi t_1 f_{\max}(\sin\theta_{\max}\sigma_x + \cos\theta_{\max}\sigma_z)\right]
	\end{equation}
	\begin{equation}
		U_{t_2} = \exp\left[-i \pi t_2 f_0\sigma_z\right]
	\end{equation}
	\begin{equation}
		U_{n}(t_1, t_2) = \left(U_{t_2} U_{t_1}\right)^n
	\end{equation}
\end{subequations}
where $f_0$ and $f_{{\max}}$ are the Larmor frequencies at the initial and maximal positions, respectively, and $\theta_{\max}$ is the polar angle of the magnetic field at $x=d_{\max}$. As before, we take the magnetic field at $x=0$ to be aligned with the $z$ axis, so that $\theta_0 = 0$. The probability of measuring the spin in the down state after $n$ shuttling cycles is therefore
\begin{equation}
	P_{\downarrow}^\mathrm{(ideal)}(n, t_1, t_2) = \abs{\bra{\downarrow} U_{n}(t_1, t_2) \ket{\uparrow}}^2.
\end{equation}
To include the effect of a finite shuttle time, we shift the waiting times as $t_1 \to t_1 + t_\mathrm{o1}$ and $t_2 \to t_2 + t_\mathrm{o2}$~\cite{unseld_baseband_2025}. We also include SPAM errors through a visibility factor $V'$ and an offset $P'_\mathrm{off}$. Because these data were acquired several days after the ones used to extract the results presented in Fig.~\ref{supp:fig:amplitude_fitting}, we do not assume the same SPAM calibration. The measured probability is then
\begin{equation}
	{P_{\downarrow}}^\mathrm{(real)}(n, t_1, t_2) = P'_\mathrm{off} + V' \abs{\bra{\downarrow} U_{n}(t_1 + t_\mathrm{o1}, t_2 + t_\mathrm{o2}) \ket{\uparrow}}^2.
	\label{eq:prob_down_nQ_real}
\end{equation}

\begin{figure}[ht!]
	\centering
	\includegraphics[width=\columnwidth]{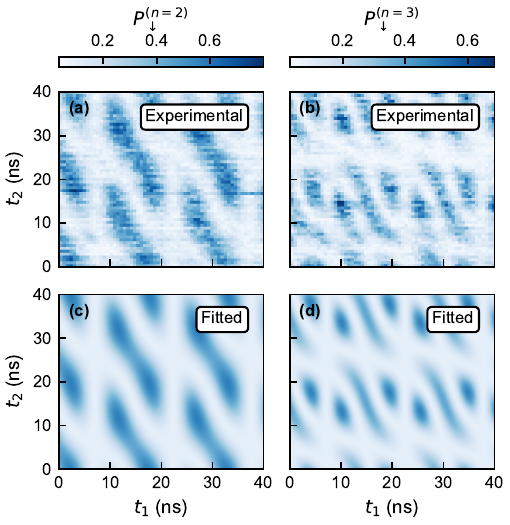}
	\caption{Repeated-shuttle single-qubit interferometry used to separate the visibility from the field tilt angle.
	Experimental probability $P_{\downarrow}$ for (a) $n=2$, and (b) $n=3$ forward-backward cycles.
	(c-d) Best-fit simulations based on Eq.~(\ref{eq:prob_down_nQ_real}).
	The consistency between the two cycle numbers constrains the visibility and the global position shift in a non-redundant way.
	}
	\label{supp:fig:fitted_pattern}
\end{figure}

In Fig.~\ref{supp:fig:fitted_pattern}(a-b), we show the experimental data for $n=2$ and $n=3$ shuttling cycles. Fitting these data with Eq.~(\ref{eq:prob_down_nQ_real}) yields the parameters listed in Table~\ref{tab:diabatic_pattern_fitting}. The extracted $f_0$, $f_{{\max}}$, $\theta_{\max}$, and time offsets are mutually consistent across the two datasets, which supports the internal consistency of the protocol. However, the fitted frequencies differ from those inferred from Fig.~\ref{supp:fig:extrapolation}a ($f_0 \sim 43$~MHz and $f_{\max} \sim 67$~MHz). Such qubit frequency differences between 2 MHz and 6 MHz for different measurements under nominally identical or nearly identical conditions have been reported previously~\cite{capannelli_tracking_2025}. We note that a 2 MHz difference corresponds to a 70 nT difference in local magnetic field. If the frequency shift is attributed to a shift in the quantum dot position of the electron, it would correspond to a displacement of about 3 nm, given the micromagnet gradient. Rather than refitting the entire field profile, we capture this drift through a global position shift $x \to x + \Delta x$, which effectively shifts the sampled magnetic field while preserving its calibrated shape.

\begin{table}[ht!]
	\centering
	\begin{tabular*}{\columnwidth}{@{\extracolsep{\fill}} c c c c c c c c}
		\hline
		$n$ & $f_0$ (MHz) & $f_{\max}$ (MHz) & $\theta_{\max}$ (deg) & $t_\mathrm{o1}$ (ns) & $t_\mathrm{o2}$ (ns) & $V'$ & $P'_\mathrm{off}$ \\
		\hline
		2 & 49.38 & 65.25 & 60.55 & 8.38 & 4.62 & 0.46 & 0.10 \\
		3 & 49.55 & 65.20 & 59.87 & 8.27 & 4.68 & 0.39 & 0.09 \\
		\hline
	\end{tabular*}
	\caption{Parameters extracted from the repeated-shuttle fits of Fig.~\ref{supp:fig:fitted_pattern}.
	The close agreement between the $n=2$ and $n=3$ results indicates that the inferred frequencies and tilt angle are not artifacts of a single dataset.
	}
	\label{tab:diabatic_pattern_fitting}
\end{table}

Treating all non-diabatic effects as effective time offsets is sufficient to extract robust Larmor frequencies, but it is too coarse to determine $\theta_{\max}$. Even if both $n=2$ and $n=3$ protocols result in a similar estimation for $\theta_{\max}$, if the shuttling speed is not sufficiently high, a fully diabatic fit yields an incorrect estimate of the tilt angle, as can be verified from a second-order Magnus expansion of the unitary evolution during the forward and backward shuttling stages. We therefore refine the analysis by computing the full-time evolution during the shuttle, using the position-dependent $B(x)$ and $\theta(x)$ reconstructed from Fig.~\ref{supp:fig:extrapolation}. Since $V'$ and $P'_\mathrm{off}$ are not expected to vary strongly with respect to the values given in Table~\ref{tab:diabatic_pattern_fitting}, the remaining free parameters are the single-qubit visibility $V$ from Eq.~(\ref{eq:prob_down_1Q_real}) and the global shift $\Delta x$. The reconstructed angle then follows from $\theta(x) = \arcsin(\sqrt{A(x)/V})$.

\begin{figure}[t!]
	\centering
	\includegraphics[width=\columnwidth]{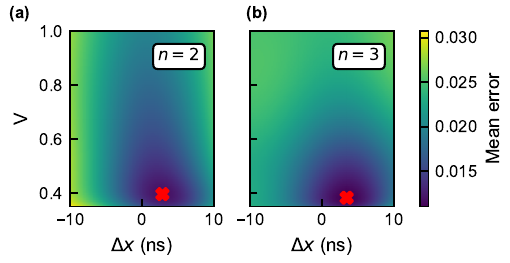}
	\caption{Mean error between the experimental data and the full-time evolution simulation of the shuttling protocol, as a function of the visibility $V$ and the position shift $\Delta x$ for (a) $n=2$ shuttling cycles and (b) $n=3$ shuttling cycles.
	The red crosses indicate the minimum error point for each case.
	}
	\label{fig:mean_error_patterns_1Q}
\end{figure}

Figure~\ref{fig:mean_error_patterns_1Q} shows the mean error between the experimental data shown in Fig.~\ref{supp:fig:fitted_pattern}~(a-b) and the full time-evolution simulation as a function of the visibility $V$, and position shift $\Delta x$ for both $n=2$ and $n=3$. The minima occur at $V = 0.40$ and $\Delta x = 2.8$~nm for $n=2$, and at $V = 0.38$ and $\Delta x = 3.4$~nm for $n=3$. These values are close to each other, indicating that the reconstructed field is not overfit to a single dataset and that the model captures the dominant coherent dynamics consistently. The average of the best-fit values for $n=2$ and $n=3$ gives a tilt angle of $\theta_{\max} =68.4\degree$. This value clearly deviates from the diabatic predictions reported in Table~\ref{tab:diabatic_pattern_fitting}, thereby demonstrating the need for a fit based on the full-time evolution. In the remainder of this section, we therefore fix $\theta_{\max}$ to the fitted value obtained above.

\begin{figure}[t!]
	\centering
	\includegraphics[width=\columnwidth]{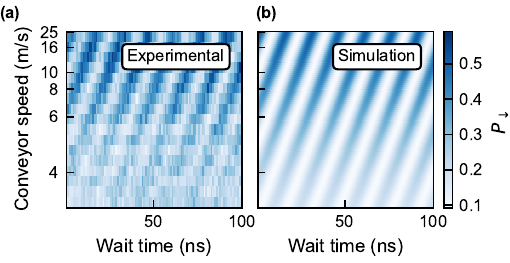}
	\caption{Predictive benchmark of the reconstructed single-qubit field profile.
	(a) Experimental probability of measuring the spin in the down state as a function of waiting time for different shuttling speeds.
	(b) Corresponding coherent simulations using the reconstructed field, without modifying its spatial profile.
	}
	\label{supp:fig:reconstructed_speed_vs_wait_time}
\end{figure}

Within the $x$-$z$-plane approximation, the single-qubit analysis therefore yields an effective reconstruction of the micromagnet field along the shuttling path, including its magnitude and polar angle. Before proceeding to the interacting problem, it is useful to benchmark this reconstruction against independent data taken at different shuttling speeds. Figure~\ref{supp:fig:reconstructed_speed_vs_wait_time}a shows the experimental probability of measuring the spin in the down state as a function of waiting time for different shuttle velocities, while Fig.~\ref{supp:fig:reconstructed_speed_vs_wait_time}b shows the corresponding simulations using the reconstructed field. The model qualitatively reproduces the main features of the data across the full speed range.

With the field profile experienced by the moving qubit constrained in this way, the remaining unknown Hamiltonian parameters are the field at the static qubit and the exchange profile. Those quantities are extracted next from two-qubit data.

\subsubsection{Two-qubit calibration}
Having fixed the field sampled by the moving qubit, we now turn to the interacting two-qubit system. The Hamiltonian is still given by Eq.~(\ref{eq:Hamiltonian_2Q}), but the magnetic field at the position of the left qubit is now fixed by the single-qubit reconstruction. The remaining unknowns are the exchange coupling $J(d)$ and the magnitude and polar angle of the field at the right qubit, $B_{R}$ and $\theta_R$, respectively.

When the exchange is weak, but non-negligible, compared to the local Zeeman energy magnitude and difference between the spins, the Larmor frequency of the left spin depends on the state of the right spin. This conditional frequency shift provides direct access to $J(d)$. In Fig.~\ref{fig:fig4}a of the main text, we show the experimental conditional Larmor frequency as a function of the conveyor distance for two initial states of the right spin: $\ket{\uparrow}$ and $\ket{\downarrow}$.

The exchange coupling can be obtained from the difference between the two conditional Larmor frequencies. 
Starting from Eq.~\eqref{eq:Hamiltonian_2Q}, we write the local magnetic fields as $\vec{B}_i = B_{i}(\sin\theta_i,0,\cos\theta_i)$, with $i=L,R$.
The Zeeman term of each spin is then diagonalized by rotating to its local eigenbasis.
Introducing the rotated Pauli operators $\sigma_{\mu,i}'$, defined through a rotation by the angle $\theta_i$ around the $y$ axis, the lab-frame operators transform as
\begin{align}
\sigma_{x,i} &= \cos\theta_i \sigma_{x,i}' + \sin\theta_i \sigma_{z,i}', \\
\sigma_{y,i} &= \sigma_{y,i}', \\
\sigma_{z,i} &= -\sin\theta_i \sigma_{x,i}' + \cos\theta_i \sigma_{z,i}'.
\end{align}
On this basis, the Zeeman contribution becomes
\begin{equation}
H'_Z = \frac{\mu_B g}{2}\left(B_{L} \sigma_{z,L}' + B_{R} \sigma_{z,R}'\right).
\end{equation}
Substituting the rotated operators into the exchange term results in the Hamiltonian
\begin{align}
H'={}& \frac{\mu_B g}{2}\left(B_{L} \sigma_{z,L}' + B_{R} \sigma_{z,R}'\right) \nonumber \\
&+ \frac{J}{4}\Big[\cos(\theta_L-\theta_R)(\sigma_{x,L}'\sigma_{x,R}'+ \sigma_{z,L}'\sigma_{z,R}') + \sigma_{y,L}'\sigma_{y,R}'\nonumber \\
&\hspace{.8cm} + \sin(\theta_L-\theta_R)\left(\sigma_{z,L}'\sigma_{x,R}'-\sigma_{x,L}'\sigma_{z,R}'\right)\Big].
\end{align}
In the far-distance regime, where the Zeeman splittings $\mu_B g B_L$ and $\mu_B g B_R$ dominate over the exchange-induced transverse couplings, the terms containing $\sigma_{x, i}'$ and $\sigma_{y, i}'$ rotate rapidly in the interaction picture.
Within the secular approximation, these transverse terms average out, leaving only the longitudinal exchange component~\cite{burkard_physical_1999,meunier_efficient_2011}.
The resulting effective Hamiltonian is
\begin{equation}
H'_\mathrm{eff} = \frac{\mu_B g}{2} \left(B_{L} \sigma_{z,L}' + B_{R} \sigma_{z,R}'\right) 
+ \frac{J}{4}\cos(\theta_L - \theta_R) \sigma_{z,L}' \sigma_{z,R}'.
\label{eq:effective_2Q_Hamiltonian}
\end{equation}

From Eq.~(\ref{eq:effective_2Q_Hamiltonian}), the left-spin Larmor frequencies are
\begin{subequations}
	\begin{align}
		f_{\uparrow} &= \frac{\mu_B g B_{L}}{h} + \frac{J \cos(\theta_L - \theta_R)}{2 h}, \\
		f_{\downarrow} &= \frac{\mu_B g B_{L}}{h} - \frac{J \cos(\theta_L - \theta_R)}{2 h},
	\end{align}
\end{subequations}
where $f_{\uparrow}$ and $f_{\downarrow}$ are the frequencies when the right spin is initialized in $\ket{\uparrow}$ and $\ket{\downarrow}$, respectively. We define their difference as $\Delta f^\mathrm{far} = f_{\uparrow} - f_{\downarrow}$. The superscript ``far'' emphasizes that this frequency splitting is extracted at separations where the qubits are still far enough apart that the protocol probes a conditional frequency shift rather than exchange oscillations. Using the above expressions, the exchange coupling is
\begin{equation}
	J = \frac{h \Delta f^\mathrm{far}}{\cos(\theta_L - \theta_R)}.
\end{equation}
At this stage, $J$ and the angle difference $\theta_L - \theta_R$ remain coupled, which is expected because both contributions modify the observed frequency splitting. The left-qubit angle $\theta_L$ is already fixed by the single-qubit calibration, so the remaining ambiguity is entirely encoded in $\theta_R$.

Resolving this remaining ambiguity requires a protocol that probes the coupled dynamics more directly. We therefore analyze a second sequence. The system is initialized in either $\ket{\uparrow \uparrow}$ or $\ket{\uparrow \downarrow}$, after which the left qubit is shuttled diabatically towards the right qubit, where the exchange is strong enough to produce oscillations in the probability of measuring the left spin in the down state $P_{L\downarrow}$. As in the single-qubit protocol, the shuttle is performed twice.
Crucially, the use of two independent waiting times provides separate control over the single-spin and exchange-induced phases. During the waiting interval $t_1$, the left spin evolves in the magnetic field at the shuttling position while acquiring a conditional phase set by the secular component of the exchange interaction, which is proportional to $J\cos(\theta_L-\theta_R)$.
By contrast, the interval $t_2$ is accumulated after the left spin has returned to its initial position, where the exchange coupling is negligible, and the evolution is governed only by the known local field of the left qubit and by the free precession of the right spin.
As a result, sweeping $t_1$ and $t_2$ produces interference features with different dependence on the exchange strength and the relative orientation of the two quantization axes. This breaks the degeneracy that would arise in a purely diabatic or single-waiting-time fit, where only the product $J\cos(\theta_L-\theta_R)$ could be inferred. The full two-dimensional pattern, therefore, allows $J$ and $\theta_R$ to be extracted independently, rather than determining only an effective conditional frequency shift.

\begin{figure}[t!]
	\centering
	\includegraphics[width=\columnwidth]{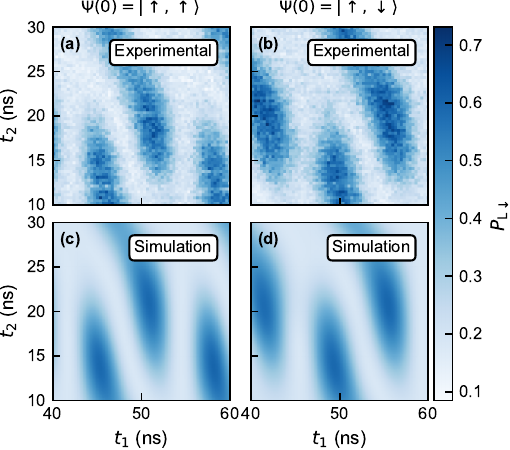}
	\caption{Large-distance two-qubit calibration in the diabatic-shuttling regime.
	(a-b) Experimental probability $P_{L\downarrow}$ as a function of the waiting times $t_1$ and $t_2$ for two initial states of the right spin.
	(c-d) Best-fit simulations obtained from the full-time evolution, using the reconstructed left-qubit field and fitting the right-qubit field together with the short-distance exchange profile.
    Here, the exchange interaction is fitted to the functional form $J(d)=J_0\exp(d/d_0)$, with fitting parameters $J_0 = 9.34\times10^{-19}$~MHz, and $d_0 = 3.55$~nm.
	}
	\label{supp:fig:fitting_diabatic_2Q}
\end{figure}

Figure~\ref{supp:fig:fitting_diabatic_2Q}(a-b) shows the experimental $P_{L\downarrow}$ as a function of the waiting times $t_1$ and $t_2$ for the two initial states of the right spin.
Using the reconstructed magnetic field from the previous section, we fit these data with the full time-evolution simulation, taking as free parameters the right-qubit Larmor frequency $f_R$, its polar angle $\theta_R$, and the exchange profile $J(d)$. For shuttling distances below $140$~nm, we extrapolate the exchange interaction using the exponential form $J(d) = J_0 \exp(d/d_0)$. Here, $d$ denotes the shuttling distance rather than the interdot separation; consequently, the exchange interaction increases exponentially with increasing $d$. This extrapolation is used only over the unmeasured short shuttle distance interval and provides the simplest monotonic continuation compatible with the measured trend. We also allow for distinct SPAM parameters $V''$ and $P''_\mathrm{off}$, as well as a global position shift $x \to x + \Delta x$.
The agreement between the experiment and simulation is good for both initial states of the right spin, showing that a single parameter set captures the conditional dynamics consistently. Table~\ref{tab:diabatic_pattern_fitting_2Q} lists the fitted parameters for the $n=2$ protocol.

\begin{table}[ht!]
	\centering
	\begin{tabular*}{\columnwidth}{@{\extracolsep{\fill}} c c c c c}
		\hline
		$f_R$ (MHz) & $\theta_\mathrm{R}$ (deg) & $V''$ & $P''_\mathrm{off}$ & $\Delta x$ (nm) \\
		\hline
		102.2 & 83.6 & 0.39 & 0.18 & 2.5 \\
		\hline
	\end{tabular*}
	\caption{Parameters extracted from the diabatic two-qubit fits of Fig.~\ref{supp:fig:fitting_diabatic_2Q}.
	These values provide the reference right-qubit field used in the gate analysis below.
	}
	\label{tab:diabatic_pattern_fitting_2Q}
\end{table}

The large interdot-distance diabatic protocol fixes the right-qubit field and polar angle, as well as the exchange interaction. To test the same Hamiltonian in a complementary limit, we next consider the regime in which the left qubit is shuttled adiabatically, with respect to the spin dynamics, to a position very close to the right qubit, left to evolve for a time $t_\mathrm{w}$, and then shuttled back to the initial position. The system is initialized in the $\ket{\uparrow \downarrow}$ state, and we measure the probabilities of finding the left and right spins in the up state, $P_{L\uparrow}$ and $P_{R\uparrow}$, as functions of $t_\mathrm{w}$ and conveyor distance.

As considered in Supplementary Information~\ref{supp:subsec:ST_osc}, a first approximation for the oscillation frequency of both qubits is given by
\begin{equation}
	f^\mathrm{close} = \sqrt{\left(\frac{J}{h}\right)^2 + \left(\frac{\Delta E_z}{h}\right)^2},
	\label{eq:frequency_close}
\end{equation}
where $\Delta E_z/h$ is the Zeeman-frequency mismatch between the two qubits. Equation~(\ref{eq:frequency_close}) is valid when the exchange coupling exceeds the Zeeman mismatch and when the difference in polar angles is small, so that $\cos(\theta_L - \theta_R) \approx 1$. Both conditions are expected near the closest approach, where the two qubits sample similar local magnetic fields. Below, we verify this assumption a posteriori.

\begin{figure}[t!]
	\centering
	\includegraphics[width=\columnwidth]{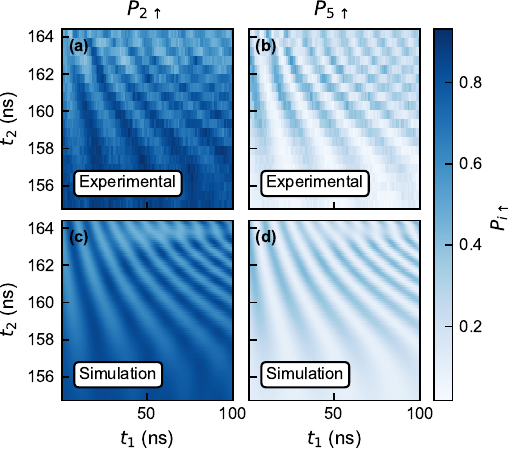}
	\caption{Close-range two-qubit calibration in the adiabatic-shuttling regime.
	(a-b) Experimental probabilities $P_{L\uparrow}$ and $P_{R\uparrow}$ as functions of waiting time $t_\mathrm{w}$ and conveyor distance.
	(c-d) Corresponding fits from the full time-evolution simulation.
	These data anchor the saturation regime of the exchange profile.
	}
	\label{supp:fig:fitting_SWAP_2Q}
\end{figure}

In fitting the data of Fig.~\ref{supp:fig:fitting_SWAP_2Q}, we again allow a global position shift with respect to the reconstructed magnetic field. We also allow a small adjustment of the right-qubit field relative to the value in Table~\ref{tab:diabatic_pattern_fitting_2Q}, since the electrostatic potential of the right dot has changed. Finally, because the two spins are measured independently, we include separate visibility factors and offsets for the two qubits, $V_L$, $P_{\mathrm{off}, L}$, $V_R$, and $P_{\mathrm{off}, R}$.

To estimate the exchange coupling $J$, we use Eq.~(\ref{eq:frequency_close}), where the frequency $f^\mathrm{close}$ is extracted from Fig.~\ref{supp:fig:ST_osc} and the Zeeman mismatch $\Delta E_z$ is determined by the fitted right-qubit Larmor frequency $f_R$. Because the exchange clearly saturates at the shortest distances, we combine the close-range estimates with the far-range data of Fig.~\ref{fig:fig4}a of the main text and fit them with the phenomenological form
\begin{equation}
	J(x) = J_\mathrm{sat} \frac{1}{1 + \exp[-(x - x_0)/\lambda]},
\end{equation}
where $J_\mathrm{sat}$ is the saturation value of the exchange coupling, $x_0$ is the distance at which it reaches half of that value, and $\lambda$ determines the steepness of the crossover. This form captures the experimentally observed transition from rapid growth at intermediate distances to saturation at closest approach.

All fitting parameters are listed in Table~\ref{tab:SWAP_pattern_fitting_2Q}. Compared with Table~\ref{tab:diabatic_pattern_fitting_2Q}, the fitted right-qubit Larmor frequency $f_R$ is slightly larger, and the polar angle $\theta_R$ is slightly smaller. This change is consistent with a modest drift in the position of the quantum dot and is accompanied by the corresponding change in the fitted shift $\Delta x$.

\begin{table}[ht!]
	\centering
	\begin{tabular*}{\columnwidth}{@{\extracolsep{\fill}} c c c c c c c}
		\hline
		$f_R$ (MHz) & $\theta_\mathrm{R}$ (deg) & $V_L$ & $P_{\mathrm{off}, L}$ & $V_R$ & $P_{\mathrm{off}, R}$ & $\Delta x$ (nm) \\
		\hline
		106.9 & 77.8 & 0.39 & 0.46 & 0.47 & 0.1 & 2.02 \\
		\hline
	\end{tabular*}
	\caption{Parameters extracted from the adiabatic two-qubit fits of Fig.~\ref{supp:fig:fitting_SWAP_2Q}.
	Separate visibilities and offsets are required because the two spins are read out independently.
	}
	\label{tab:SWAP_pattern_fitting_2Q}
\end{table}

In the regime of Fig.~\ref{supp:fig:fitting_SWAP_2Q}, the left-qubit angle is nearly constant, $\theta_L \approx 71^\circ$, while the fitted right-qubit angle is $\theta_R \approx 78^\circ$. The difference is therefore only about $7^\circ$, so that $\cos(\theta_L - \theta_R) \approx 0.99$. Equation~(\ref{eq:frequency_close}) thus incurs only a one-percent correction from the angle mismatch. The final fit gives $J_\mathrm{sat}/h = 166.4$~MHz, $x_0 = 163.3$~nm, and $\lambda = 3.23$~nm, capturing both the saturation of the exchange at short distances and its rapid decay at larger separations. Moreover, $J_\mathrm{sat}/h$ is larger than the Zeeman mismatch $f_R - f_L \sim 33$~MHz, which further justifies the use of Eq.~(\ref{eq:frequency_close}) at closest approach.

At this point, the field profile of the moving qubit, the effective field at the static qubit, and the exchange law are all constrained by the experiment. The calibrated Hamiltonian can therefore be used predictively to ask which nonlocal operations are available under the experimentally relevant shuttling conditions.

\subsubsection{Predicted Two-Qubit Gate Set}
We now use the calibrated Hamiltonian to assess which two-qubit gates the model predicts are accessible with the shuttling protocol. Because the magnetic field shows some dataset-to-dataset variability, we fix the global shift to $\Delta x = 2$~nm and the right-qubit Larmor frequency to $f_R = 102.2$~MHz, i.e., the values extracted from the diabatic two-qubit calibration. We first consider the regime of small quantization-axis mismatch, $\theta_R = 83.6^\circ$, and then examine how the results depend on that choice.
In order to focus on the available non-local two-qubit gates and trace out possible single-qubit gates, we use the method described in Ref.~\cite{fernandez-fernandez_spin-orbit-enabled_2025}. Then, available gates should be interpreted as gates that can be performed directly in a single forward and backward shuttling round, up to possible single-qubit gates on the two spin qubits before and after shuttling.
As an example, in this analysis, we consider that $\mathrm{CZ}\sim \mathrm{CNOT}$, making no distinction between the two gates.

\begin{figure}[t!]
	\centering
	\includegraphics[width=\columnwidth]{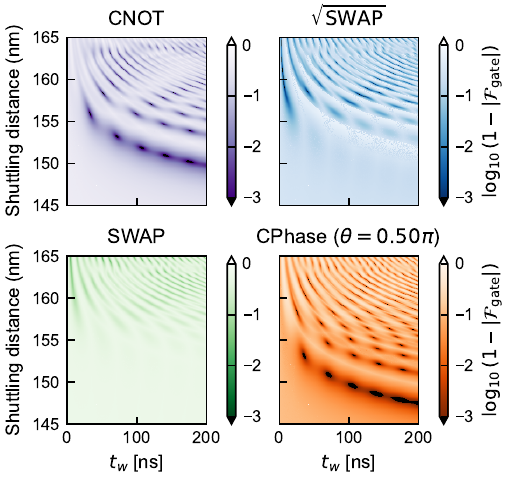}
	\caption{Model-predicted infidelity maps for representative two-qubit gates, indicated in the panel titles, as functions of conveyor distance and turning-point waiting time $t_w$.
	Each panel uses its own color scale; in all cases, darker colors correspond to lower infidelity.
	The calculations use the calibrated field and exchange profiles with $f_R = 102.2$~MHz, $\theta_R = 83.6^\circ$, and $\Delta x = 2$~nm.
	}
	\label{supp:fig:2Q_gates_examples}
\end{figure}

The shuttle velocity is fixed at $v = 16.2$~m/s, while the maximum shuttling distance is varied between $x_{\max} = 145$~nm and $x_{\max} = 165$~nm, and the waiting time at the turning point is between $t_\mathrm{w} = 0$~ns and $t_\mathrm{w} = 200$~ns. Figure~\ref{supp:fig:2Q_gates_examples} shows the model-predicted infidelity of several representative two-qubit gates as a function of conveyor distance and turning-point waiting time. The maximum fidelities exceed $99.99\%$ for CNOT, $\sqrt{\mathrm{SWAP}}$, and CPhase$(\pi/2)$, whereas the SWAP gate fidelity reaches about $97\%$. These values should be interpreted as an upper bound because of systematic errors in the case of fully coherent dynamics, rather than as directly measurable gate fidelities, including noise and other imperfections. The broad low-infidelity regions for CNOT and CPhase$(\pi/2)$ indicate robustness against modest timing and positioning errors, which is encouraging from an experimental perspective. By contrast, the narrow bands associated with $\sqrt{\mathrm{SWAP}}$ imply that these gates require substantially tighter calibration.

To visualize the reachable gate set more globally, we can plot in the Weyl chamber all gates generated by these shuttling protocols with infidelity below $10^{-3}$, as shown in Fig.~\ref{fig:fig4}f of the main text. At this threshold, the protocol covers approximately $\mathcal{V} \sim 66.8\%$ of the Weyl chamber. Here $\mathcal{V}$ denotes the covered volume fraction of a Weyl chamber used in the numerical analysis. The figure should therefore be read as a visualization of a three-dimensional coverage metric. A notable exception is the segment connecting the $i$SWAP and SWAP gates, which does not reach below the $10^{-3}$ threshold. Relaxing the threshold to $10^{-2}$ makes much of this line accessible as well, except for the SWAP point itself, and increases the total coverage to approximately $\mathcal{V} \sim 97.3\%$.

\begin{figure}[!tbh]
	\centering
	\includegraphics[width=\columnwidth]{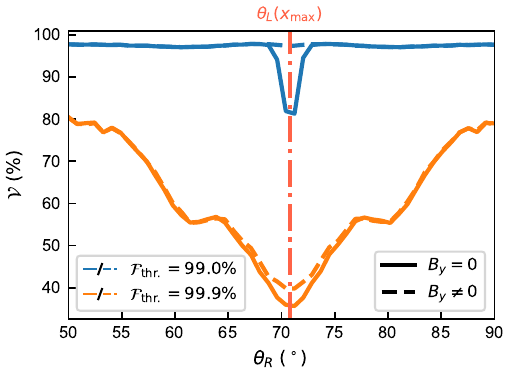}
	\caption{Predicted Weyl-chamber coverage as a function of the right-qubit polar angle $\theta_R$ for infidelity thresholds $10^{-2}$ (blue) and $10^{-3}$ (orange).
	Solid lines correspond to the $x$-$z$-plane model ($B_y = 0$), whereas dashed lines include a finite $y$ component through the azimuthal perturbation discussed in the text.
	The vertical dot-dashed line marks $\theta_R = \theta_L(x_{\max}) = 70.8^\circ$, where the two qubits have nearly aligned quantization axes and the predicted coverage is minimal.
	}
	\label{supp:fig:coverage_swap}
\end{figure}

Because the right-qubit tilt angle is not directly measured, it is important to quantify how the reachable gate set depends on $\theta_R$. The results are shown in Fig.~\ref{supp:fig:coverage_swap}, where we plot the Weyl-chamber coverage as a function of $\theta_R$ for two infidelity thresholds, $10^{-2}$ and $10^{-3}$. The coverage is smallest near $\theta_R = \theta_L(x_{\max})$, where $\theta_L(x_{\max})$ is the polar angle of the field at the left qubit for the largest conveyor distance used in the experiments. In that regime, both qubits share nearly the same quantization axis, and the interaction generates a less diverse set of nonlocal gates, which explains the reduced coverage.

The present data do not determine whether the magnetic field has a finite $y$ component. To test the robustness of the conclusions, we therefore repeat the coverage analysis with a small azimuthal variation $\phi$ that grows linearly with $x$, namely $\phi(x) = 10^\circ \times (x / x_R)$, where $x_R = 200$~nm is the position of the right qubit. This ansatz is not a fit; it is a representative perturbation used to gauge sensitivity to $B_y \neq 0$. The corresponding results are shown by the dashed lines in Fig.~\ref{supp:fig:coverage_swap}. A finite relative azimuth enlarges the reachable set for the looser threshold $10^{-2}$, but produces no substantial change at $10^{-3}$. The high-fidelity conclusions are therefore robust against a small magnetic-field component outside the $x$-$z$-plane. Overall, within the experimentally motivated parameter range, the Weyl-chamber coverage lies between roughly $36\%$ in the most conservative scenario and $80\%$ in the most favorable one.
Within this calibrated model, the shuttling protocol is predicted to support a broad family of high-fidelity entangling gates and to cover a substantial fraction of the Weyl chamber, even under conservative assumptions about the right-qubit field orientation and a possible out-of-plane $(B_y)$ magnetic-field component.

\end{document}